 \documentclass[sigconf]{acmart}
\AtBeginDocument{%
  \providecommand\BibTeX{{%
    \normalfont B\kern-0.5em{\scshape i\kern-0.25em b}\kern-0.8em\TeX}}}

\acmConference[Conference acronym 'XX]{Make sure to enter the correct
  conference title from your rights confirmation emai}{June 03--05,
  2018}{Woodstock, NY}
\usepackage{colortbl}
\usepackage{multirow} 
\usepackage{color} 
\usepackage{wrapfig}

\usepackage[symbol]{footmisc}

\usepackage{hyperref,tablefootnote,footnotehyper}

\newcommand{\yh}[1]{{\color{black} {#1}}}

\acmSubmissionID{2765}

\begin{document}

\title{\sysName: Creating Interactive Scenes Using Modular LLM and Graphical Control
}

\author{Hui Ye}
\affiliation{%
  \institution{Hong Kong University of Science and Technology}
  \city{Hong Kong}
  \country{China}}
\email{huiyehy@outlook.com}

\author{Chufeng Xiao}
\affiliation{%
  \institution{Hong Kong University of Science and
Technology}
  \city{Hong Kong}
  \country{China}}
\email{chufengxiao@outlook.com}

\author{Jiaye Leng}
\affiliation{%
  \institution{City University of Hong Kong}
  \city{Hong Kong}
  \country{China}}
\email{jiayeleng2-c@my.cityu.edu.hk}

\author{Pengfei Xu}
\affiliation{%
  \institution{Shenzhen University}
  \city{Shenzhen}
  \state{Guangdong}
  \country{China}}
\email{xupengfei.cg@gmail.com}

\author{Hongbo Fu}
\authornote{Corresponding author.}
\affiliation{%
   \institution{Hong Kong University of Science and Technology}
   \city{Hong Kong}
   \country{China}}
\email{hongbofu@ust.hk}

\renewcommand{\shortauthors}{Hui Ye, Chufeng Xiao, Jiaye Leng, Pengfei Xu, Hongbo Fu}

\newcommand{\sysName}[0]{\emph{MoGraphGPT}}

\begin{abstract} 
Creating interactive 
scenes {often} involves complex programming tasks. Although large language models (LLMs) like ChatGPT can generate code 
from natural language, {their} 
output is often error-prone, particularly when scripting {interactions}
among multiple elements. The linear conversational structure limits {the editing of individual elements,}
{and lacking graphical and precise control}
complicates {visual integration.}
To address these issues, {we \yh{integrate
an \emph{element-level modularization}} technique that processes textual descriptions for individual elements through separate LLM modules, with a central module managing interactions {among elements.}}
{This modular approach allows for refining each element independently.} We design a graphical user interface, {\sysName}, which combines modular LLMs with enhanced graphical control to generate code{s} for {2D} interactive scenes. {It enables direct integration of graphical information and offers quick, precise control through automatically generated sliders.}
Our comparative evaluation against \yh{an AI coding tool, {Cursor Composer}, as the baseline system} 
{and}
a usability study show \sysName~significantly improves {easiness,}
controllability, and refinement in creating complex 2D {interactive scenes} 
with multiple visual elements {in a coding-free manner}.

\end{abstract}



\begin{teaserfigure}
  \hspace{1mm}
  \includegraphics[width=0.95\textwidth]{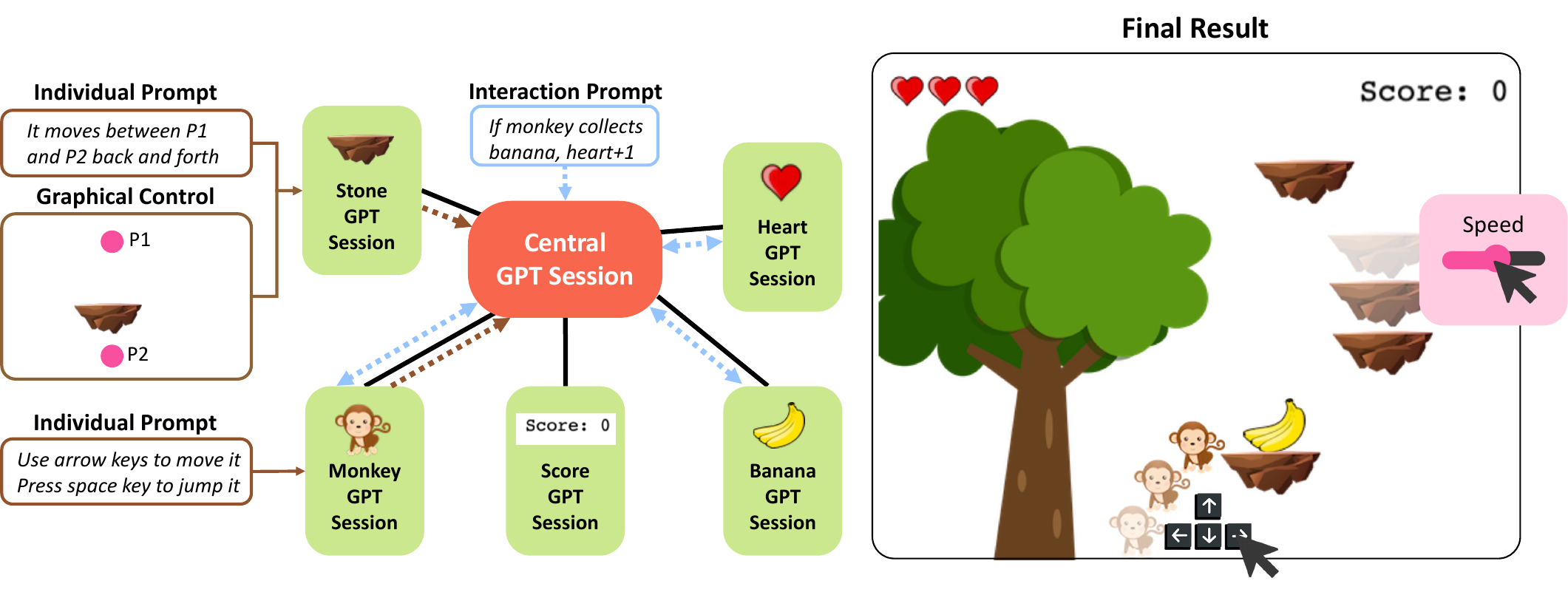}
  \caption{\yh{We introduce~\sysName~to facilitate the easy creation of interactive scenes using a graphical user interface powered with modular LLMs. 
  Users can input the text descriptions for individual element properties or behaviors, or interactions {among} 
  multiple elements, {along with the directly specified or drawn graphical information 
  }into our system. The code for each element and interactions among elements {is} 
  generated from 
  different modules, with 
  automatically-generated sliders to control effects precisely.
  }
  }
  \label{fig:teaser}
\end{teaserfigure}

\maketitle
\section{Introduction}
{Interactive scenes combine} 
artistic visuals with interactive elements, allowing users to immerse themselves in a richly crafted environment{. They} 
are widely used in various scenarios, including video games \cite{supermario, tetris, zelda}, educational tools \cite{khanacademy, scratch}, and interactive demonstrations \cite{phET, exploratorium}. {Creating} 
interactive scenes {often} involves coding with engines like Unity \cite{unity}, or frameworks like Phaser \cite{phaser}, requiring skills in scripting interactivity, managing animations, and debugging. This complexity 
is crucial for delivering engaging user experiences {but can be difficult for beginners} \yh{\cite{myers2008designers,kazi2014kitty, zhang2020flowmatic, yang2024gpt, sweetser2024large}}.

Recently, Large Language Models (LLMs) like ChatGPT \cite{openai_chatgpt} have emerged as powerful {tools} 
for users, enabling them to generate code {given} 
natural text as input. {They} 
can assist users in writing scripts \cite{de2024llmr}, creating {simple} animations \cite{lan2023application}, and even debugging \cite{zhang2023critical}, significantly reducing the time and effort required for coding tasks \cite{tian2023chatgpt}. Some ad-hoc tools for code generation and refinement \yh{\cite{github_copilot,jupyter, codepen, glitch, claud_artifacts}} have also been proposed to provide real-time suggestions. 

However, {directly using LLMs to generate code for interactive scenes}
may have four main {issues:}
1) \textbf{code quality} -- 
{LLM-based code generation approaches}
may produce incomplete or incorrect code sometimes \cite{li2022competition,nijkamp2022codegen,roziere2023code}. {They} 
often require a clear context to generate relevant code, which may be challenging in complex projects. {Generating} 
interactive scenes 
usually requires scripting how elements in {a} scene 
interact with each other and how users {interact} 
with the scene. The logic and relationship can become complex, especially when multiple elements are involved. {The generated code tends to be error-prone, so users need to provide extra text input to refine it.}
2) \textbf{lack of {editing} independency} -- the linear conversational nature of LLMs limits independent editing of individual elements\yh{, especially for non-expert users \cite{zamfirescu2023johnny}}. 
The model{s} may forget previous interactions, leading to disjointed responses, {especially when the conversation is long and involves many {scene} elements}. \yh{The modular code generation and refinement in emerging AI tools often require understanding project and code structures for hard modularization, otherwise the updated results may intertwine codes across other files to produce unintended modification.}
3) \textbf{lack of graphical control} -- 
{it is difficult} to directly integrate graphical information into the text input. {Users} need to manually estimate the {exact} 
graphical information (e.g., position, size, path, region) and translate it into textual input, which is not intuitive and direct \cite{masson2024directgpt}. 
{4) \textbf{lack of precise control} -- fine modifications on the generated effects (e.g., effect parameters) require users to adjust text prompts iteratively \yh{\cite{dang2022prompt}}, which usually traverse users between excessive and inadequate controlling.}

By reviewing videos on using LLMs to create interactive scenes and comparing existing tools on code generation using content analysis \cite{harwood2003overview}, we identified the challenges of applying existing tools for generating codes for 2D dynamic and interactive scenes. Based on the insights, we design and develop \sysName, an interactive system for creating interactive scenes using modular LLM and graphical control without coding. To enable independent code generation for individual {scene} elements, 
we \yh{integrate an \emph{element-level modularization}}
{technique}, which {maintains} 
independent LLM modules for {individual elements}, 
where users can input text descriptions to generate class codes for the element features and actions. On top of individual modules, a central LLM module manages the relationships and interactions of all the elements. 
To avoid {the central module's lack of} context-specific details, we guide LLM to {distill}
contextual information about the variables and functions along with the generated code for individual elements. So the central LLM can generate correct and desired interaction code based on the contextual information. We implement this concept in the \sysName~system, a graphical interface for users {to create} 
2D interactive scenes using modular LLM and graphical control. In \sysName, users draw/import their prepared elements or ask the system to generate elements and then input the text for each element and multiple-element interaction separately. Four types of graphical proxy, including point, line, curve, and region, can be specified by direct pointing and drawing, and then 
be explicitly mentioned in the text prompt. The positions, \yh{orientations},
and sizes of elements can be adjusted manually and 
integrated {into} the generated code {automatically}. Our system automatically generates sliders from the code for users {to interactively adjust the} 
parameters of the effects in the code, {to reduce the text description input}. \yh{{A comparative study}
shows 
that \sysName~outperforms \yh{{a state-of-the-art} 
AI coding tool, Cursor Composer}} 
in \yh{creating interactive scenes}
for fixed tasks{. An open-ended study demonstrates \sysName} {enables users to} 
create diverse 2D games, animations, and demonstrations easily.

This work makes the following main contributions:
\begin{itemize}
\item{A content analysis
of video tutorials on creating interactive scenes using ChatGPT and existing AI tools for code generation.}
\item{\yh{The integration of \emph{element-level modularization}}
for controlling independent code generation for individual elements in individual LLM modules, as well as interaction generation and overall management for all elements based on element context in the central module.}
\item{A graphical interface allowing users to create interactive scenes by inputting text prompts to modular LLMs, integrating graphical information into prompts directly, and iteratively adjusting generation results and parameters using {additional text inputs} 
and sliders precisely.}
\item{Two evaluations that compare the performance of {\sysName} and \yh{Cursor Composer} and {validate} the system usability.
}
\end{itemize}

\section{Related Work}
\label{sec:related_work}
\subsection{LLM-based Code Generation and Improvement}
LLM-based code generation leverages advanced language models \cite{li2022competition,nijkamp2022codegen,roziere2023code} to translate natural language prompts into executable code. Although these models are powerful at streamlining software development and enhancing productivity, many issues arise from the generated code, including compilation and syntax errors, wrong outputs, maintainability problems, not following standard coding practice, need refactoring, security smells, etc \cite{tian2023chatgpt, liu2024refining,liu2024no,siddiq2024quality}. In particular, ChatGPT struggles to generate code for new and unseen problems \cite{tian2023chatgpt}. Lengthy prompts might have negative impacts on the code generation \cite{tian2023chatgpt, liu2024refining}. {These issues} 
could be due to the increased {code complexity for more complex problems}, 
making it harder for {LLM models} 
to generate a correct and complete solution \cite{liu2024refining}. 

Researchers have proposed \yh{to use Chain-of-Thought \cite{wu2022ai, wei2022chain}} 
to make LLMs more controllable for complex tasks,  
{especially} for text-based organization tasks. For coding tasks, many decomposition strategies are employed using interactive decomposition \cite{huang2024anpl}, block-based hierarchical structure \cite{ritschel2022can}, node-based diagrams \cite{wu2022promptchainer}. 
CoLadder \cite{yen2023coladder} further enables programmers to decompose tasks flexibly. These works mainly focus on general programming tasks, which typically involve static code generation and logic implementation. \yh{Modularized LLM generation mechanisms have been researched by Tree-of-Thoughts and its variations \cite{yao2024tree, besta2024graph}. \yh{They focus on general thought exploration, and we aim for a specific scenario of interactive scene creation. We integrate textual and graphical inputs with LLM code generation to enable intuitive control over element behavior and interaction.
Agentic workflows \cite{qian2024chatdev} achieve agent-level modularization and communication for entire software development across roles (Table \ref{tab:comparison}). In contrast, 
we emphasize element-level modular creation. We focus on precise and independent control over code generation for individual elements and their interaction.}
Kim et al. \cite{kim2023cells} explore a design framework for users to control the modularized generation of configuration components for writing tasks. We extend this framework to a specific application of interactive scene creation, where elements act as objects. This shift presents unique challenges in managing element interactions. We address this by encapsulating the code from the interaction generation or refinement for each element within its module while invoking the interaction logic in a central module.
}

\begin{table*}[]
\caption{\yh{The comparison among the closely related works.
}}
\footnotesize
\begin{tabular}{|c|c|c|c|c|c|c|}
\hline
{ }                     & { \textbf{Task}}                   & { \textbf{Method}} & { \textbf{\begin{tabular}[c]{@{}c@{}}Multiple Element \\ Interaction\end{tabular}}} & { \textbf{\begin{tabular}[c]{@{}c@{}}Code \\ Modularization\end{tabular}}} & { \textbf{\begin{tabular}[c]{@{}c@{}}Graphical \\ Control\end{tabular}}} & { \textbf{\begin{tabular}[c]{@{}c@{}}Precise \\ Control\end{tabular}}} \\ \hline
{ \textbf{MoGraphGPT}}  & { Interacitve scene creation}   & { LLM-based}       & { Yes}                                                                              & { Element-level}                                                           & { Yes}                                                                   & { Yes}                                                                 \\ \hline
{ \textbf{DrawTalking \cite{rosenberg2024drawtalking}}} & { Interacitve scene creation}   & { Rule-based}      & { Yes}                                                                              & { /}                                                                       & { Yes}                                                                   & { No}                                                                  \\ \hline
{ \textbf{Spellburst \cite{angert2023spellburst}}}  & { Generative art creation}         & { LLM-based}       & { No}                                                                               & { Node-level}                                                              & { No}                                                                    & { Yes}                                                                 \\ \hline
{ \textbf{DirectGPT \cite{masson2024directgpt}}}   & { Text/code/vector images edition} & { LLM-based}       & { No}                                                                               & { No}                                                                      & { Yes}                                                                   & { No}                                                                  \\ \hline
{ \textbf{ChatDev \cite{qian2024chatdev}}}     & { General software development}    & { LLM-based}       & { Yes}                                                                              & { Agent-level}                                                             & { No}                                                                    & { No}                                                                  \\ \hline
{ \textbf{Cursor \cite{cursor2023}}} & { General programming}             & { LLM-based}       & { Yes}                                                                              & { Code line/block/file-level}                                              & { No}                                                                    & { No}                                                                  \\ \hline
\end{tabular}
\label{tab:comparison}
\end{table*}

\subsection{LLM-powered Tools for Visual Design and Development}
LLMs {have been} 
employed to enhance visual design \cite{brade2023promptify} and development processes {and generate} 
2D static visual design{s}, 
such as personalized logos \cite{xiao2024typedance}, interior color designs \cite{hou2024c2ideas}, \yh{storybook \cite{yan2023xcreation}}, and editorial illustrations \cite{liu2022opal}. Designing dynamic 2D visuals {with LLMs}, however, presents greater challenges. Unlike static designs, animations require consideration of timing, motion, and interaction. Keyframer \cite{tseng2024keyframer} leverages LLMs to empower users in creating animations by generating keyframes based on textual descriptions. LogoMotion \cite{liu2024logomotion} develops an LLM-based system that automatically generates content-aware animations for logos by synthesizing code from visual layouts. Spellburst \cite{angert2023spellburst} introduces a node-based interface that enables users to explore creative coding through natural language prompts for interactive visual design \yh{with precise parameter control (Table \ref{tab:comparison})}. \yh{While these works enhance the use of LLMs in animation and dynamic visual design, they do not address interactions between multiple elements, thus overlooking the challenges of independent control. \yh{v0 \cite{v0_vercel} can generate interactive and dynamic effects, but lacks clear control for individual elements and graphical information.}}


Games are one of the most popular applications of interactive scenes. Researchers have explored the potential of LLMs in game content design, including investigating the use of video game description language \cite{hu2024generating}, automated level design and generation from text \cite{sudhakaran2024mariogpt, todd2023level}, and co-creative game design \cite{anjum2024ink}. They aim to enhance the game content {design}
process, making it more accessible for creators to {design} 
game experiences. 
{Differently, we provide an LLM-based solution, allowing users to interactively create complete games without coding.}

Some studies further explore the potential of LLMs in application development, including using communicative agents to support software development \cite{qian2024chatdev}, assisting end-users in generating robot programs \cite{bimbatti2023can}, enabling zero-code generation of trigger-action IoT programs \cite{li2023chatiot}, and providing an AI-augmented system for autonomous visual programming learning for children \cite{chen2024chatscratch}. They aim to lower the barriers to programming, making it easier for non-experts to create and manage complex applications. {Compared to them, our work focuses on a novel system for creating 2D interactive scenes.}

\subsection{Supporting Creating 2D Interactive Scenes}
The traditional practice of creating 2D interactive scenes requires artists to create frame-by-frame animations and integrate them with programming to enable user interactions and responses. To simplify the production process, previous works introduce sketch-based interactions to animate virtual elements \cite{liu2020posetween}, manipulate kinetic 
textures
\cite{kazi2014draco}, and using filters for dynamic illustrations \cite{xing2016energy}. Besides dynamic effects, Kitty \cite{kazi2014kitty} enhances the sketching interface for interactive illustrations, which {involve} 
more user interactions. These works are powerful for creating diverse, fascinating, dynamic effects, but expect users to have drawing skills 
{for specifying animation effects}. Recently, several tools employ visual programming interfaces using blocks \cite{ye2024prointerar}, node-graphs \cite{snap_lensstudio, unreal_blueprints}, and flowcharts \cite{chen2021entanglevr, yigitbas2023end, zhang2020flowmatic} to build interactive scenes. {For example,} as a pioneering platform, Scratch \cite{scratch} enables users to create animations and games through a block-based coding environment. {Our graphical interface is largely inspired by the interface of Scratch. However, instead of explicit coding in Scratch, our system uses LLMs to generate code given natural language text inputs, aiming to lower the barriers to creating interactive scenes.} 


DrawTalking \cite{rosenberg2024drawtalking} is closely related to our work and enables {the creation of} interactive worlds using sketching and speaking. The main difference between DrawTalking and our work is that we introduce LLMs to generate codes for scenes automatically \yh{(Table \ref{tab:comparison})},  \yh{while their dependency-tree structure is difficult to replace directly 
with an LLM model, a point confirmed by the authors of DrawTalking}. The rule-based interactions in DrawTalking can involve multiple effects, but {lack} 
flexibility and generalization for scripting customized effects. 
{For example,
{our system supports the use of} 
different {development} frameworks to better create scenes according to users' intention, e.g., Phaser to create games, p5.js to create creative coding effects. {Such a feature is more difficult} 
to achieve in DrawTalking since it is designed with defined rules.  
In addition, the natural language input in DrawTalking needs to conform to the rules while ours allows users to describe {desired} 
scenes freely due to the understanding capability of ChatGPT.}

\vspace{-1.1mm}
\subsection{\yh{Interacting with LLM from Input and Output}}
Recently, researchers have explored \yh{interacting with}
LLM input and output to enable more controllability. For example, DirectGPT \cite{masson2024directgpt} offers an intuitive interface for users to engage directly with {LLMs}, 
making it easier to input prompts with direct manipulation. \yh{It focuses on a {task different from ours}—editing text, code, and vector images—so it does not address dynamic interactions among multiple elements or allow for precise adjustment of editing parameters (Table \ref{tab:comparison}).} Graphologue \cite{jiang2023graphologue} enhances this interaction by allowing users to explore LLM responses through interactive diagrams, which help visualize the relationships and structures within generated content, thus clarifying complex ideas. Meanwhile, Visual ChatGPT \cite{wu2023visual} combines conversational capabilities with visual foundation models, enabling users to talk, draw, and edit visuals simultaneously, creating a richer and more dynamic engagement experience. ChainForge \cite{arawjo2023chainforge} serves as a visual toolkit for prompt engineering and hypothesis testing, allowing users to iteratively refine their inputs and analyze outputs in an intuitive visual format, thereby enhancing the overall interaction with LLMs. Compared to these works, we {use graphical control and direct manipulation to specify both input and output}.
For input, we integrate graphical information {(}by specifying or drawing visual proxies{)} into text prompts. For output, we allow users to quickly refine the effect parameters via sliders and value inputs.

\section{Formative Steps}

To better understand the challenges of utilizing LLMs to generate codes for interactive scenes, we employed a {mixed-method} 
approach in our formative steps: (1) content analysis on {video tutorials} 
about using ChatGPT to generate codes for building interactive scenes; (2) further analysis {of} 
existing AI coding tools; 

\subsection{Content Analysis on {Video Tutorials} 
about Creating Interactive Scenes Using ChatGPT}
\label{sec:content_analysis}
\yh{Due to the lack} of well-developed workflows and criteria for creating interactive scenes using LLMs, it is not easy to understand the existing challenges according to \yh{experts' experiences.} 
So we resort to online videos -- many users uploaded 
video tutorials {documenting their} 
experience in using LLMs to create games or make interactive demos for popular video platforms (e.g., {``Can AI code Flappy Bird? Watch ChatGPT try''}
\footnote[3]{\url{https://www.youtube.com/watch?v=8y7GRYaYYQg}}). We want to distill valuable insights from their videos.

\textbf{\yh{Corpus and Methodology.}} We searched {for} the videos on popular video websites (e.g., YouTube, Vimeo, TikTok) using keywords such as ``GPT for interactive scenes'', ``GPT for games'', ``GPT for animations'', and ``GPT for dynamic effects''. Through the first round of searching, we collected 208 video clips. After a thorough filtering process, we retained 56 videos that specifically guided or demonstrated how to use GPT to build a complete interactive or dynamic scene. \yh{All the videos are in English and feature a single speaker. The styles include full-screen screencasts, screen recordings with annotations, tutorial formats, and picture-in-picture {screencasts}. 
The duration of the videos ranges from 4 minutes 13 seconds to 26 minutes 48 seconds. The programming languages covered in these videos include C\#, JavaScript, Python, GML, Lua, and Scratch pseudocode.} Two of our authors employed the open-coding approach \cite{charmaz2008constructionism} to analyze the content of the selected videos. \yh{We began by watching each video multiple times to {understand its content comprehensively.}  
In the initial coding phase, we identified explicit difficulties mentioned in the videos, as well as challenges reflected in the creation process. We focused on common issues faced by {those} users when using GPT for creating interactive scenes, {their} strategies 
to overcome these difficulties, and the problems that persisted even after applying these methods. We developed a coding framework that categorized the identified difficulties and strategies into several themes.}
\yh{We then compared individual results and summarized the findings into overarching themes.} The coding process was iterative, allowing for refinement as new insights emerged.



\textbf{Findings.} We distilled three main {issues} as follows. 

\emph{Independent generation and refinement \yh{(29/56)}}. 
{Three} types of strategies are mainly used for generating codes: (1) describing all the elements in the scene at one time and then iteratively adjusting the results; (2) describing each element and element interactions step by step and manually adjusting the code snippet of different elements;
\yh{(3) describing the scene and asking ChatGPT to implement a basic version as the start, and then {adding} 
more features iteratively.} The first strategy does not require much programming understanding, but it \yh{may produce incorrect results.} 
Refining one element would sometimes affect other elements. 
For example, in {reproducing the} 
Super Mario game, adding a moving feature to one platform might also make another static platform move. This is due to ambiguous references and a limited understanding of ChatGPT. The second \yh{and third strategies} require coding skills to some degree. Sometimes GPT generates incorrect results, so the users need to use their prior knowledge to fix the errors. The dependent results will also require manual adjustment and differentiation.

\emph{Graphical control \yh{(42/56)}.}
To enable GPT to generate graphical scene interfaces or demos, there are {three} types of strategies to employ: (1) let GPT generate {graphic effects} using simple descriptions 
and then adjusting {them} 
using text prompts or manual coding refinement; (2) preparing a 2D snapshot of a 
desired graphical scene {with} 
accurate graphical information of each element; {(3) copy and paste the generated code to game engines like Unity and manipulate elements}. The first strategy {often} produces random results -- even different for every trial. Users need to use wording like ``make the rectangle larger'' and ``move it to the top'' to adjust the accurate properties. If users would like to make a random game for fun, {this strategy could} 
be acceptable. {Otherwise, they have to bear a tedious adjustment process}. 
The second strategy requires users to make a lot of preparation effort and then translate this graphical information into text. Some information, like user-defined curved paths, is very difficult to describe in text. So, they only use a rough representation of the results they create. {In the third strategy, users need a good understanding of both coding and game engines. For those without coding skills, filling the gap between the generated code and the manipulation in game engines requires GPT to guide users step by step to find the correspondence.}

\emph{Precise refinement \yh{(27/56)}.} Once the scene code is generated, users may refine the specific effects, such as the moving speed and the rotation radius of elements. 
In ChatGPT, they input the text again using comparative expressions (e.g., make the Mario jumps lower each time the space key is pressed, let the star move slower). However, they usually do not have an intuitive understanding of the magnitude of the parameters. So, they need to refine it back and forth to achieve the desired effect.

{In summary,} the analysis identified three main challenges in creating interactive scenes using ChatGPT: the difficulty of independent generation and refinement of code, the lack of graphical control requiring extensive manual adjustments, and the need for precise parameter adjustments due to users' limited understanding of effect magnitudes. These challenges highlight the complexities users face when leveraging LLMs 
for interactive scene creation.

\subsection{Further Analysis on Existing AI Coding Tools}
\textbf{Methodology.} We collected and reviewed six common AI coding tools (i.e., GitHub Copilot, Cursor, Tabnine, Codeium, Replit Ghostwriter, and Amazon CodeWhisperer). For the analysis of these tools, we mainly focus on the {three distilled} issues 
in Section \ref{sec:content_analysis}. 

\textbf{Findings.}
These tools provide functions for generating code snippets or blocks, {either} 
independently or considering the entire file context. However, \yh{implementing individual elements in an interactive scene, such as those in game development, often requires creating entire classes.}

The challenge lies in constructing these classes and maintaining the interactions 
between multiple classes. While these tools can generate basic class structures, they typically lack support for managing complex interactions, such as communication between a player character class and an enemy class. {Therefore,} 
users often need to manually refine and adjust the generated code to ensure that the components work together effectively, {demanding} 
a solid understanding of object-oriented programming principles.
\yh{Some tools, like Cursor \cite{cursor2023}, support modular code generation and {refinement} 
by selecting project files as context to constrain the generation. However, it {performs modular code generation in a soft manner}. 
{In other words}, 
it treats them as contexts, which can lead to unintended modifications across other files. \yh{Such confusion can intertwine interaction code and individual behavior code, causing inconsistent updates and chaotic modifications.} For users seeking more controllable hard modularization, a solid understanding of the entire project structure and code framework is essential {with Cursor}, creating a significant barrier to entry.
}

Since these tools are primarily designed for general programming tasks, none of them supports graphical control of the scripted elements, making it difficult to create interactive scenes directly and intuitively. This limitation forces users to rely solely on text-based refinement, which can be cumbersome when managing complex visual components. Without a graphical interface, users cannot easily manipulate or visualize elements in real-time, leading to a tedious trial-and-error process. The precise control also relies heavily on text-based adjustment. While these tools can provide suggestions for modifying {element} properties, 
they lack the ability for direct manipulation of graphical elements. This means users must translate their visual intentions into code, which can be a time-consuming process, especially for intricate designs. They may spend significant time fine-tuning parameters through trial and error rather than simply dragging and dropping elements or adjusting them visually. 


\subsection{Design Considerations}
\label{sec:design_consideration}
Based on the above findings, we envision an ideal LLM-based tool for creating interactive scenes should consider the following points:

D1. Independent {code} generation and control on elements: refining individual elements does not affect others.

D2. Context-aware code generation: independency will not lose context.

D3. Graphical control: 
integrating graphical information directly into text prompts.

D4. Easy and precise parameter control: direct manipulation of effect parameters.

\begin{figure*}[t]
\includegraphics[width=0.8\linewidth]{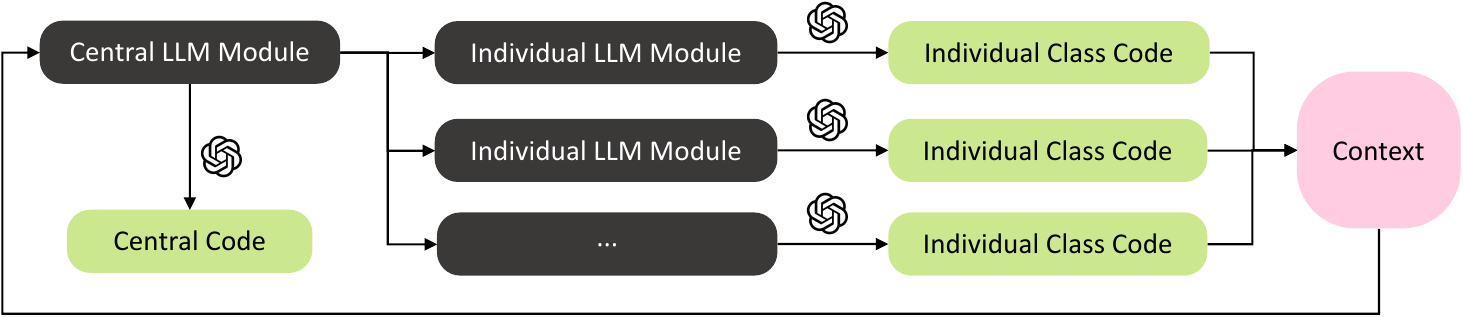}
  \caption{\yh{The framework of our context-aware LLM modularization technique. The central LLM module generates and maintains central code. It manages individual LLM modules to generate individual class codes. The contextual information is extracted from individual codes and input to the central LLM module for reference. }
  }
  \label{fig:framework}
\end{figure*}

\section{MoGraphGPT System}


According to the design considerations, we first \yh{integrate an element-level} context-aware \emph{modularization} technique (D1, D2) to help generate code for individual elements and interactions for multiple elements {(Section \ref{sec:LLM_modularization})}. We further design and develop a graphical interface, \sysName, combining modular LLMs with graphical control for users to create 2D interactive scenes {(Section \ref{sec:UI})}. It enables direct integration of graphical information (D3) and offers quick, precise control through automatically generated sliders (D4).

\subsection{\yh{Element-level} Context-aware LLM Modularization}\label{sec:LLM_modularization}
2D interactive scenes contain elements in various forms. We define the \emph{element} as {a} 
general representation of the content within these scenes, \yh{encompassing both individual visual components and broader concepts}. For example, \yh{a layered character animation includes animations for individual body parts as well as a global transformation, meaning both the parts and the entire body are considered {elements}  in our design}. 

\yh{Our element-level}
context-aware \emph{modularization} technique \yh{(Figure \ref{fig:framework})} 
opens modular LLMs for individual elements and {uses} a central LLM module {to manage} 
interactions and relationships {among} 
elements. \yh{It employs a hierarchical 
structure where the central module oversees coordination, while the individual modules operate independently. This design ensures a clear and cohesive update logic, allowing modifications to a single element without affecting others. When creating interactions, the relevant function code is updated within the element class, while the central module uniformly calls these functions.} {We employ ChatGPT-4o Mini as our LLM {model} 
in our implementation.}

\textbf{Individual LLM Modules for Individual Elements.} Each individual element in a 2D interactive scene is associated with its own LLM module. These individual modules are {used}
to 
generate and maintain 
class {codes}
{for} 
their respective elements. For example, when creating a Super Mario platform game, the Mario element has its own LLM module (Figure \ref{fig:framework}), which generates a class named Mario for its own properties (e.g., sizes) and behaviors (e.g., using arrow keys to control its movement) from the text input. {To} 
modify Mario's properties and behaviors {with additional text prompts}, it will search the created Mario's module and continue to update there. {This} 
approach allows each element to operate independently, enabling users to customize and enhance each element without disrupting other elements with rapid iteration and testing.

\begin{figure*}[t]
\includegraphics[width=0.95\linewidth]{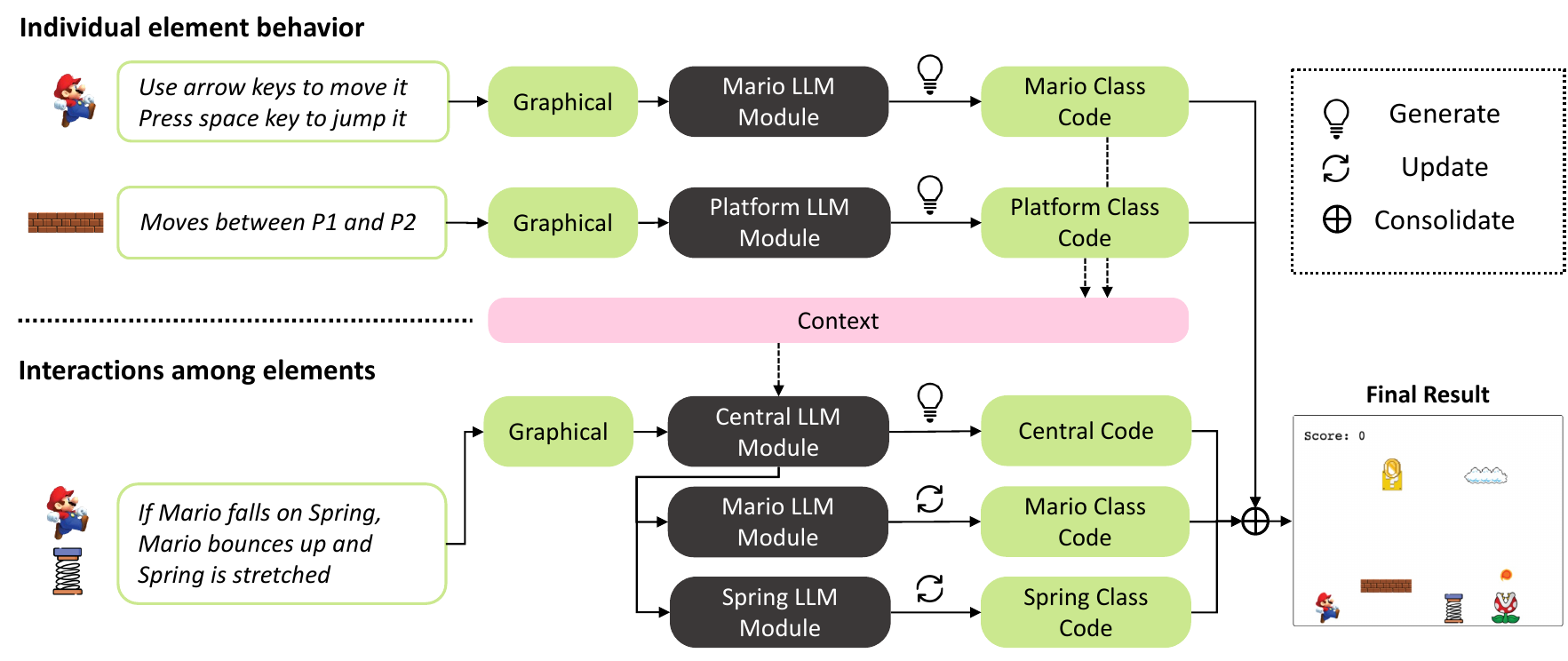}
  \caption{
  \yh{\sysName~workflow. When users input text prompts for individual elements, our system integrates graphical information into prompts and sends {them} to individual modules to generate class codes {(Top)}. For interactions {(Bottom)}, prompts {with the integrated graphical information} go to the central LLM, which creates the central code. It then notifies individual LLM modules to update their codes with new variables and functions. Changes are reflected in real-time, and the central and individual codes together form the final result.}
  }
  \label{fig:workflow}
  \vspace{-2mm}
\end{figure*}

\begin{figure*}[t]
\includegraphics[width=0.99\linewidth]{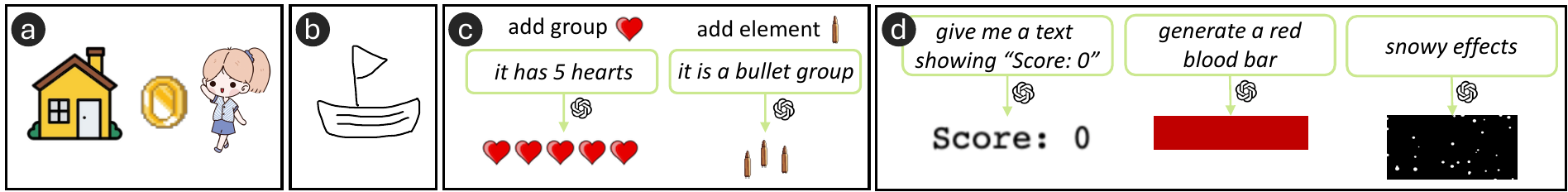}
  \caption{
  \yh{Four ways to create elements in our system. (a) Upload an image. (b) Draw a sketch. (c) Add {a} group 
  and let LLM generate {a group of elements (with a user-uploaded element image)}, 
  either explicitly mentioning ``group'' in text prompt or not. (d) Ask LLM to generate elements.}}
  \label{fig:elemen_type}
\end{figure*}

\textbf{Central LLM Module.}
In contrast to the individual LLM modules, the central LLM module serves as the orchestrator of interactions and relationships {among} 
elements (Figure \ref{fig:framework}). It is responsible for instantiating classes from individual modules, coordinating their communication, {and} managing interactions among elements, {thus} ensuring that they work together cohesively within the interactive scene. For example, in the Super Mario platform game  (Figure \ref{fig:workflow}), the central module generates codes for instantiating all the elements and scripting interactions among elements (e.g., when Mario falls on the spring, Mario bounces up and the spring is stretched). Importantly, when generating interaction code, it may involve variables and behaviors specific to individual elements. To prevent interference between elements, we instruct it to 
define the code of variables and functions for each element 
within their respective classes {(e.g., Mario bouncing code in Mario class, spring stretching code in spring class)} and to call these functions in the central module. This approach allows the central module to directly invoke functions from individual modules while keeping their definitions separate. As a result, modifications to individual elements do not impact the interaction code, maintaining the integrity and functionality of the overall system.

\textbf{Contextual Communication between Modules.}
Independent code generation will lead to a lack of contextual information. To address this issue, 
we design a contextual communication mechanism  (Figure \ref{fig:framework}) between the central module and individual modules. Each time {the} code for {an} individual element is generated, we guide LLM to also provide a summarized overview of the class, including the class name, variables (name, initial value, and short description), and functions (name, argument, return value, and short description). {Please refer to the supplementary materials for more details.} Such information is then compiled into a context information repository. When generating the code from the central module, this {context} 
is referenced, enabling the central module to maintain an understanding of the overall state of all elements in the scene. It can directly access the variables 
and call the functions defined in the element classes. If a user revises any element {class}, 
both {its} 
code and context information will be updated accordingly. Additionally, if the central module modifies or updates variables and behaviors for elements, this will also be reflected in the contextual information. This dynamic updating ensures that the central module remains aware of all changes, promoting a more responsive and flexible operation.

By integrating the strengths of the individual and central LLM modules with contextual communication, our context-aware modularization technique not only enhances independence in code generation but also fosters a more dynamic and interconnected interaction creation. 

\subsection{Graphical Interface}\label{sec:UI}
\begin{figure*}[t]
\includegraphics[width=0.99\linewidth]{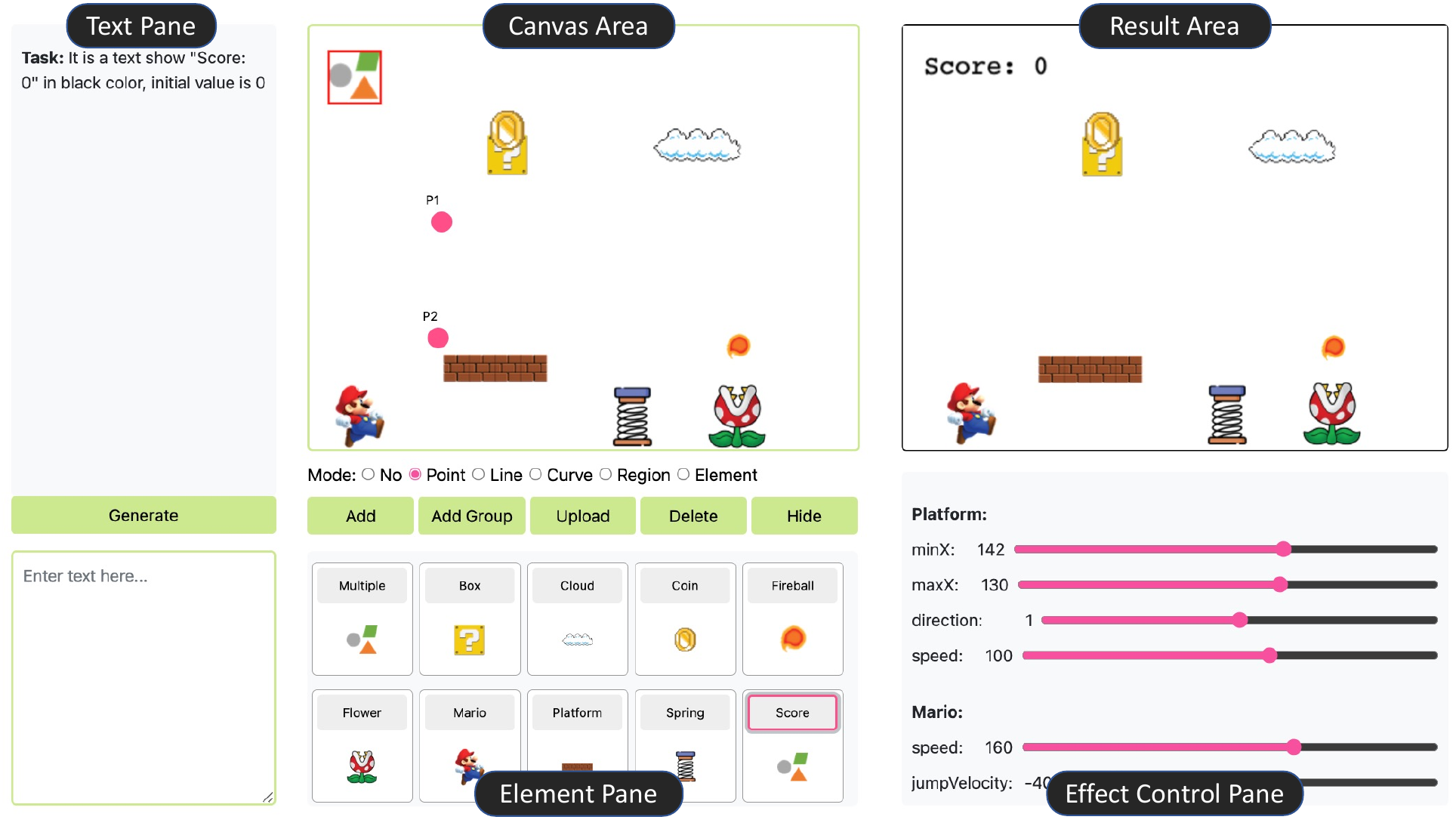}
  \caption{\yh{\sysName~user interface. Element Pane contains the buttons and preview images for all the created elements in the scene. Canvas Area shows all the elements that can be manipulated by users directly. Once users press the ``Generate'' button, the result is generated or updated in the Result Area. Effect Control Pane displays the automatically generated parameter values and sliders for precise control.}}

  \label{fig:UI}
\end{figure*}

In our graphical interface \sysName~  (Figure \ref{fig:UI}), context-aware LLM modularization and graphical control are seamlessly integrated to facilitate the creation of 2D interactive scenes using natural language inputs and graphical specifications.
\yh{Our target users are \yh{those with no or limited programming skills,}
and our goal is to help them create interactive scenes rather than learning programming. To simplify the user experience and avoid overwhelming newcomers, we do not reveal {the generated} code 
in the UI, as 
common in other tools \cite{scratch,python_playground,flutter}. Instead, we focus entirely on prompts and graphical elements, encouraging users to engage with this specialized tool for scene creation rather than transitioning to full programming.

}

\begin{figure*}[t]
\includegraphics[width=0.99\linewidth]{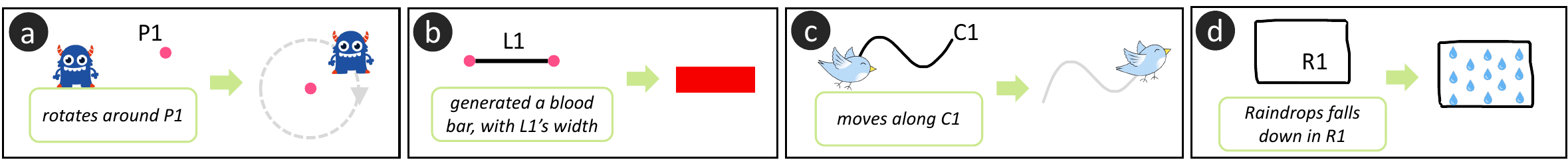}
  \caption{
  {We allow users to specify four types of graphical inputs: (a) point, (b) line, (c) curve, and (d) region. {Users can refer to their names in the text prompts}. 
  }} 
  \vspace{-2mm}
  \label{fig:spatial}
\end{figure*}

\textbf{Element Creation.}
{Users have the flexibility to upload, draw, and request our system to generate elements for them.
Users can press the ``Upload'' button to upload an image element (Figure \ref{fig:elemen_type}(a)) and draw elements with 2D sketches on the canvas area (Figure \ref{fig:elemen_type}(b)). If users have not prepared any images, they can create an empty asset by pressing the ``Add'' button and then input text descriptions to ask the {associated} individual module to generate an element for them (Figure \ref{fig:elemen_type}(d)), such as texts, graphics, and particle effects. Besides single elements, users can create element groups in two ways: 1) the user can press the ``Add Group'' button and upload an element image, and then add a text description to let our system generate a group of elements {with the uploaded image}; 2) 
the user can upload an image element and let our system generate an element group with a proper text prompt including words like ``group'' (Figure \ref{fig:elemen_type}(c)).
}



Once the element is created, it is displayed in the canvas area (Figure \ref{fig:UI}) and rendered in the result area (Figure \ref{fig:UI}). It opens a dedicated ChatGPT session for that element in the left text pane (Figure \ref{fig:UI}). {Our system automatically switches the GPT session to an element after its selection (by pressing its associated button or clicking on it in the canvas)}. 
The first element in the element pane serves as a central proxy, representing the central session. By clicking on this proxy, users can access the central session in the left pane.

\textbf{Graphical Control.}
Once {an} 
element is created, the user 
can move, rotate, and scale it on the canvas. These graphical properties are updated in real-time in the generated code, as displayed in the result area. Since describing graphical properties in a natural language can be challenging, we {introduce} a drawing mode allowing 
users to specify four types of graphical inputs: point, line, curve, and region (Figure \ref{fig:spatial}). They can switch to a certain mode and draw on the canvas. After completing their drawings, each input is labeled with an index, designated as \( \mathit{P}_i \), \( \mathit{L}_i \), \( \mathit{C}_i \), \( \mathit{R}_i \), respectively. Users can then reference these labels explicitly in their text input, facilitating {explicit} 
communication of their graphical specifications.

\textbf{Text Input.}
{We} allow users to input any text to describe the interactive scenes. For properties and behaviors of individual elements, users enter the text in the module of each element by selecting element button and press the ``Generate'' button. Then the code for the element is generated and rendered in the result area. For interactions among multiple elements, users input text in the central module by selecting ``Multiple'' button and press the ``Generate'' button to send their request. Since each element has its own ChatGPT session, users do not need to mention the element names explicitly in the individual sessions. Instead, users can use pronouns such as ``it'', ``each of them'', or ``all of them'' to refer to specific elements.

\textbf{Precise Refinement.} After the code for each element is generated, we let LLM to extract the defined variables and their current values. Then the system automatically generates sliders and number input fields in the effect control pane (Figure \ref{fig:UI}), with the range normalized. This allows users to quickly and precisely adjust the parameter values (e.g., movement speed, shake amplitude) without needing to describe the desired changes in text and ask for refinement again. 

{
\textbf{Result Testing.} Users can watch and test the created results in the result area (Figure \ref{fig:UI}) at any time during the creation process. Any change, including text revision, slider revision, element manipulation, 
in the canvas area will lead to instant updates in the result area.

}

\section{Implementation}
\sysName~is built using JavaScript for the front-end client, and our back-end uses OpenAI’s ChatGPT \cite{openai_chatgpt} API, specifically using the GPT-4o Mini model.

\subsection{Prompt Design}

\textbf{Code Template Generation.} We first ask GPT to generate a template class code with a basic setting according to the JavaScript framework that users {will use}. 
The framework is automatically {determined} 
for users when {they input} 
general descriptions in {the} central module. For example, if the user inputs ``I want to make {a} 
platform game'', {the} Phaser framework will be {selected;}
if he/she inputs ``make 
a creative coding project'', the p5.js framework is initialized. When a new element is created, an initial class template code will be added to the scene folder, and the central.js is also automatically initialized with the elements instantiated.

\textbf{Code Generation for Individual Elements and Multiple Elements.} {To guide GPT to generate output in a controlled manner, the text prompt has to follow a certain format. For individual element code,} we design the prompt {format} 
mainly including \emph{Task}, \emph{Requirement}, \emph{Reference Code Template}, and \emph{Output Format}. The ``elementname'', ``effect'', and ``framework'' in the format will be replaced by the created element names, user's input texts (including the translated graphical information), and suggested framework name. Then they forms a text prompt to GPT{, which generates the} 
code and context for {each} element. For interactions codes among multiple elements, we design the prompt in similar format with individuals, except for an extra context information and the output format. Please refer to the supplementary materials for the specific prompt designs. 


\subsection{Code Integration}
Once receiving the response, for individual module{s}, we extract the code, insert it to the element classes{, and insert} 
the context to the context information repository. When updating {an} element later, the newly generated code will replace the original one, and the newly updated context will be compared to the original ones and then {updated}. 
For the central module, we copy the code generated and replace the original {one}, 
and insert the newly defined variables and functions in the relevant individual element class code. The context information is also compared and updated correspondingly. The insert positions is determined by the pre-defined or guided-maintained flags (e.g., ``//variable start'', ``//variable end'', ``//function start'', ``//function end'') in the code.

\yh{
\section{Comparative Study}

To evaluate the effectiveness of element-level modularization, graphical control, and precise refinement of our system, we designed a \emph{within-subject} study {to compare} 
\sysName~with a state-of-the-art tool. We selected Cursor \cite{cursor2023} as the baseline since it is a well-recognized AI code editor that allows for partial code modification and simultaneous updates of files or modules. Participants created interactive effects using both \sysName~and Cursor.}

\yh{
\subsection{Baseline Setup}
The Cursor Composer with the GPT-4o Mini Model served as our baseline tool. It supports both full and partial code generation and refinement from text input, applicable to one or multiple files. However, it lacks graphical control features. To ensure a fair comparison, we prepared a basic code template identical to that of the \sysName~system, featuring a central JavaScript file along with separate JavaScript files for individual elements. Participants {could} 
select {those} files as context to enhance modular code refinement. To observe the results in real time, we launched a live server that rendered the outputs immediately. Participants were instructed to focus solely on text input, context selection, and result rendering, without visibility into the underlying code.

\subsection{Participants and Apparatus}
}
\yh{We invited 10 participants (aged 22-34, M: 28, SD: 3.77, 6 females and 4 males, U1-U10) from our personal and university network. They include{d} 4 university students and 6 staff. They {had} 
diverse backgrounds, including atmospheric environment (U1, U10), fine arts (U4), interaction design (U8), computer science (U2, U5-6), entrepreneurship (U3), chemistry (U7), and business (U9). On a self-rated 5-point scale (1-no to 5-strong) for coding experience, 2  (U4 and U9) of them rated 1, 3 users (U3, U8, U10) rated 2, 1 user (U7) rated 3, 3 users (U1-2, U5) rated 4, and 1 user (U6) rated 5. In a self-rated 5-point scale (1-no to 5-strong) for ChatGPT using experience, 2 users (U9-10) of them rated 2, 3 users (U1, U7-8) rated 3, 3 users (U3-5) rated 4, and 2 users (U2, U6) rated 5. They used ChatGPT for searching information (U1, U4),  polishing writing (U1-3, U6-7, U9-10), checking codes (U2, U6-7), generating codes (U2, U5). The study was conducted on a laptop running both systems, and the participants could use a keyboard, touchpad, and mouse for inputting.}


\begin{figure*}[t]
\includegraphics[width=0.99\linewidth]{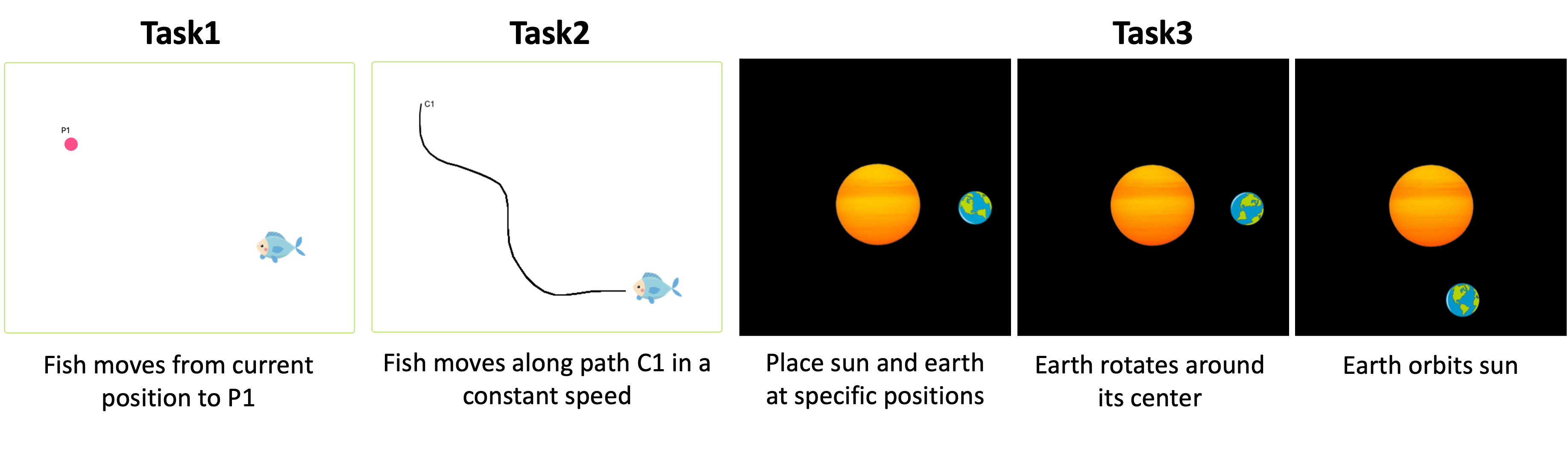}
  \caption{Three tasks in the comparative study.} 
  \label{fig:fix}
\end{figure*}

\subsection{Tasks} 
We designed three tasks (Figure \ref{fig:fix}) for users to reproduce the following effects using \sysName~and \yh{Cursor Composer}: (Task1) a fish moves from a specific point to another point; (Task2) a fish moves along a curved path with a constant speed; (Task3) a three-step iterative animation: place the sun and the earth {at specific positions} 
on the canvas, {then} let the earth {rotate around its own center}, 
{and finally} let the earth {orbit} 
the sun while keeping self-rotation. These three tasks are common in 2D interactive scenes and 
involve typical features such as behaviors of single elements and interactions between two elements, spatial properties like positions, translations, rotations, paths, and speed. Users were required to create their results as similar as the given effect example video. In particular, the following features should be similar to the target effects as much as possible: (Task1) the positions of starting point and ending point in the canvas; (Task2) the moving path and the speed; (Task3) the positions of the sun and earth and the rotation speeds {of the earth}.

\subsection{Procedure} 
We gave participants a \yh{15-minute} introduction of the tasks and two systems and allow them to try the systems freely. Then, we showed them both an image and a video for the target effect of {each task}, 
and they can further see the image and video during the whole study process. To avoid the learning effects of our system, we asked each participant to first use \sysName~and then \yh{Cursor Composer} to reproduce the target effect for each task. Once the participants considered that their created target effect has been reproduced successfully, it was double-checked by two of our authors. If both 
of us reach 
a consensus, it was considered a complete result. He or she can move to the next step. If the participant tried over  \yh{10 minutes} 
for 
similar text prompts {but} 
the system still does not provide a clear result, or the participant thinks the effect is very difficult to achieve and he/she does not have any idea for it, it is considered a incomplete result, and it can move to the next step as well. After completing all the tasks, they were asked to fill in the questionnaire on a 5-Likert scale. The questions are {elaborated} 
in Figure \ref{fig:sub-compare}. We then conducted semi-structured interviews with them to collect their feedback, including the differences between \sysName~and \yh{Cursor Composer} and our observations during the study. We recorded the time spent on each task, text prompts, operations, and results. The whole process was audio-recorded and later transcribed with their agreement by filling out an informed consent form. In compensation for their time, each participant received a 13-USD gifted card for about \yh{one-hour} participation.

\subsection{Data Analysis} 
\begin{figure*}[t]
\includegraphics[width=0.99\linewidth]{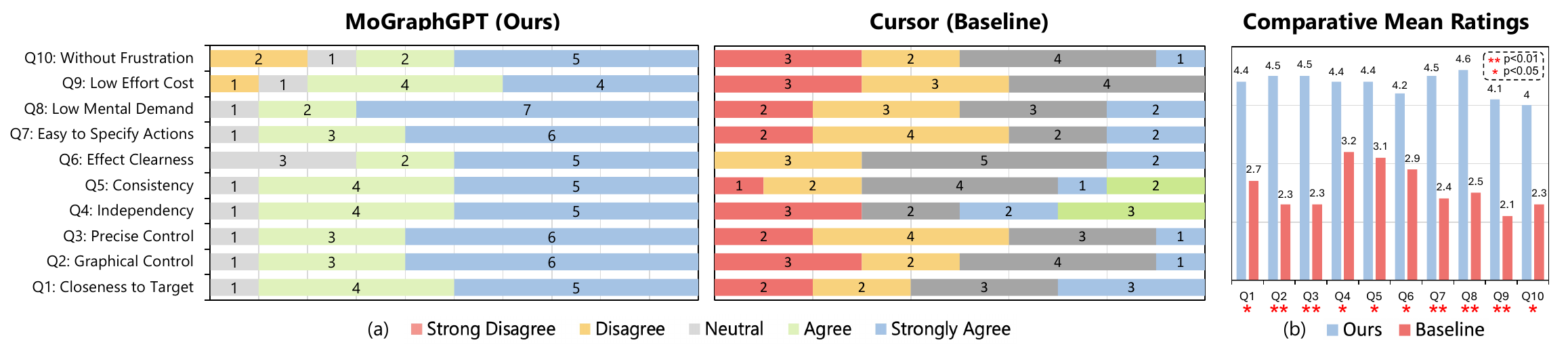}
  \caption{Subjective ratings on \sysName~and \yh{Cursor Composer.} 
  For the scores, the higher, the better.}
  \label{fig:sub-compare}
\end{figure*}

The {questionnaire} includes personal information background questions, subjective ratings (Closeness to target, Graphical control, Precise refinement, Effect independency, Effect consistency, Effect clearness, Easy to specify action, Mental demand for formulating prompts), and selected questions from NASA-TLX (Figure \ref{fig:sub-compare}). Objective metrics consist of time taken, the number of prompts, and prompt word counts. We conducted Wilcoxon signed-rank tests to analyze significant differences.

\begin{table}[]
    \centering
    \caption{The performance
of MoGraphGPT and Cursor across three metrics averaged over 10 participants.
Note that the time, prompt number, and prompt length are averaged across participants for the total of the three tasks.}
    \label{tab:stat_comparison}
    \begin{tabular}{l|c|c|c}
    \hline
                           & Time (s)   & Prompt \textit{N} &  Prompt \textit{L}  \\ \hline
    Ours                     & 402.40      & 4.80          & 27.70         \\ \hline
    \yh{Cursor}                      & 1339.00  & 17.20         & 269.30        \\ \hline
    Reduction & 69.57\%  & 69.34\%       & 89.21\%       \\ \hline
    \end{tabular}
\end{table}



\subsection{Results}
\textbf{Completion.} 
\yh{All participants successfully completed {each} task in our system within 6 minutes.
However, U7 {spent over 10 minutes on Task1 and U4-6 and U8-10 spent over 10 minutes on Tasks} 
using Cursor but were still unhappy with their results. They found Cursor was hard to handle graphical information, such as specific positions and curved paths. Despite attempts to change descriptions or correct responses, Cursor often retained the original results or produced undesirable effects. 
{Table \ref{tab:stat_comparison} compares the performance of \sysName~with Cursor across three metrics averaged over the 10 participants {for the three tasks}: total completion time {(in seconds)}, prompt number, and prompt length (in words) for all the tasks. We also calculated the reduction rates for these metrics by averaging individual improvements of all the participants with \sysName~compared to Cursor. The results indicate \sysName~achieves desired outcomes in significantly less time and with fewer prompts than Cursor.}}

\textbf{Time.} 
\yh{
The average time spent on each task using \sysName~is significantly lower than Cursor{, as confirmed by the Wilcoxon signed-rank test (p<0.01)}: Task1: \yh{M: 46.1s (SD: 12.5s) and M: 268.0s (SD: 134.4s) for \sysName~and Cursor, respectively; 
Task2: M: 152.1s (SD: 73.9s) for \sysName~and M: 448.0s (SD: 110.7s) for Cursor; Task3: M: 204.2s (SD: 79.6s) for \sysName~and M: 623.0s (SD: 161.6s) for Cursor.} {Here, we trim the time for cases over 10 minutes to 10 minutes.} 
Our graphical specification feature enables participants to quickly define positions, moving paths, and both absolute and relative positions quickly. Participants (U1, U4, U6-7, U9-10) were able to adjust motion parameters, such as speed, using sliders and numerical inputs with precision. In contrast, participants using Cursor spent considerable time formulating prompts to integrate graphical information. Some (U1, U4-6, U8) struggled with precise descriptions for several minutes, especially when Cursor continued to produce undesired results. Participants (U1-3, U5, U7-9) often had to try multiple word variations to adjust motion parameters.
In Cursor, adding context does not guarantee independent control, often requiring participants to attempt multiple times and impose additional constraints. In \sysName, individual and interaction behaviors can be created in the earth and central sessions just one trial. 
This is because our modular structure provides clear division when updating interactions--defining function code for elements within their individual classes and invoking these functions in the central module. In contrast, Cursor’s context offers only soft constraints, causing interaction code and individual behavior code to become intertwined, leading to inconsistent updates and chaotic modifications.}

\textbf{Prompt numbers and length.}
\yh{Our system results in significantly fewer and shorter prompts, as confirmed by Wilcoxon signed-rank tests (p < 0.01). In Task1 and Task2, the participants used pronouns to refer to each element in element sessions and mentioned the created graphical proxies (e.g., P1, C1) in their prompts. In contrast, accurate mention of element names and graphical details is required in Cursor. 

Participants described the relative positions to the canvas (e.g., left-top, right-bottom, U1-3, U5, U7-9), estimated specific coordinates (U4, U6, U10), and refined results  using reference objects and iterations (e.g.,  set it higher to the corner, make them much closer, U1-2, U4-6, U8-9). They adjusted coordinates through multiple iterations and articulated the curved motion path with terms like "curves with two circles" (U2), "waved curves" (U7), and "tilde" (U1, U3, U9). They further specified the shape with phrases such as "make it more curved" (U2) and "let it curve to the left-bottom and then top-right" (U3).

In Task3, even for the second step, 
Cursor often failed to achieve target effects on the first trial. Success typically came only after participants added the earth file as context and retried multiple times. Some participants (U1-2, U5, U8-10) directly inputted text for the third step by instructing the earth to orbit the sun, resulting in the earth orbiting but losing its self-rotation effect. This necessitated rewriting the prompt to include this effect, such as ``let the earth rotate around the sun while also rotating around itself.''  In contrast, with \sysName, participants could input the self-rotation instruction in the earth module and the orbiting effect in the central module independently, requiring less prompt engineering effort.
}  

\yh{
\textbf{Subjective ratings and qualitative feedback.}
We analyzed the participants' rating and feedback, distilling our findings into four key aspects.

\emph{User Overall Experience.} 
Overall, subjective ratings and feedback indicate that our system offers a more intuitive, easy-to-use, and effective experience 
compared to Cursor.
As shown in Figure \ref{fig:sub-compare}, \sysName~significantly outperformed Cursor across all metrics. {Wilcoxon signed-rank tests confirmed the significance of all aspects (p < 0.05). As shown in Figure \ref{fig:sub-compare} (b),} the participants had a strong preference for \sysName, {particularly in terms of {Q2 - Graphical Control, Q3 - Precise Control, Q7 - Easy to Specify Actions, Q8 - Low Mental Demand, and Q9 - Low Effort Cost (p<0.01).}} 

\emph{Strength of Modularization.} Participants (U2, U4-6, U7, U9) highlighted the modularization of our system as a key advantage over Cursor (Q4 \& Q5). For example, U4 noted, ``the refinement effect for a single element [in our system] is clear (Q6), while in the other tool [Cursor], the modular refinement is ambiguous due to a lack of clear distinction between different elements.''

\emph{Strength of Graphical Control.} Participants appreciated our GUI for its intuitive manipulation of visual elements (Q2). Several participants (U1, U4, U6, U9) were surprised that they could create target spatial effects in just a few seconds using our system (Q7). Even those who redrew curves in Task 2 to better match the target path (U2, U8-9) expressed a willingness to experiment without significant effort (Q8 \& Q9). Most participants noted that relying solely on text to describe shapes made it difficult and cumbersome for Cursor to generate accurate results, leading to frustration (Q10).

\emph{Strength of Precise Control.} The precise control over effect parameters (Q3) in \sysName~was praised by participants. They highlighted its intuitiveness and effectiveness, even for the interaction effects with multiple elements (e.g., orbit radius, orbit speed). In contrast, they found that using text to adjust parameters in Cursor ``does not have a reference'' (U7) and required balancing between excessive and inadequate control (U2, U9).
}


\begin{figure*}[t]
\includegraphics[width=0.99\linewidth]{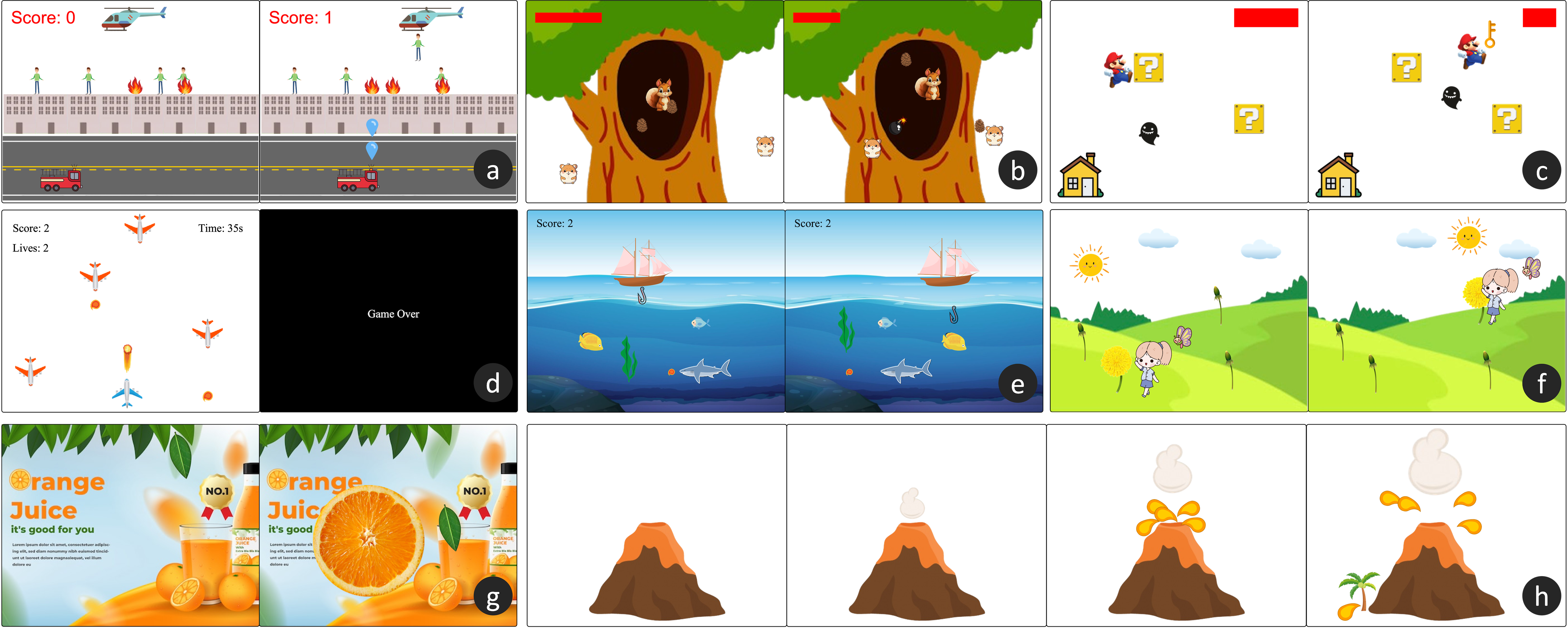}
  \caption{\yh{A gallery of selected results in the open-ended study. (a) Two-player Rescue Game (P2). (b) Squirrel Guard Game (P1). (c) Scavenger Hunt Game (P4). (d) Airplane War Game (P4). (e) Sea Fishing Game (P6). (f) Interactive Animation Demo (P3). (g) Website Ad Design Demo (P5). (h) Dynamic Illustration for Academic Paper (P6). Please refer to supplementary materials for detailed descriptions.}}
  \label{fig:result}
\end{figure*}

\section{Open-Ended Study}
To further evaluate the usability and expressiveness of \sysName, we invited participants for an open-ended study, allowing them to create their own desired 2D interactive scenes freely.

\textbf{Participants and Apparatus.}
We recruited 6 participants (aged 25-33, M: 29.5, SD: 2.66, 4 females and 2 males, P1-6), and \yh{three of them} 
participated in our comparative study. 
The study was 
conducted on 
a laptop or a tablet running \sysName, and the participants could 
use a keyboard, touchpad, {stylus}, 
and mouse for inputting and drawing.

\textbf{Procedure.}
Before the study, the participants were 
asked to think about 
interactive scenes that they would like 
to create. This process mainly allowed us to prepare the element images for {those elements requiring image uploading  
from our devices}. At the beginning of the study, after a 
brief introduction and guidance of our system, the participants 
started creating using \sysName. We stayed next to them, 
answering questions and providing verbal guidance whenever they had doubts. After they finished the creation, they played or showed the created results for us to demonstrate the final scenes.

\textbf{Results.} The 
participants created 10 results in total. Figure \ref{fig:result} shows parts of result snapshots, including 5 games (1 two-player game (Figure \ref{fig:result}(a)) and 5 single-player games (Figure \ref{fig:result}(b)-(e))), 1 interactive animation demo (Figure \ref{fig:result}(f)), 
1 website ad interaction design demo (Figure \ref{fig:result}(g)), and 1 dynamic illustration for an academic paper (Figure \ref{fig:result}(h)). Each result contains 4-8 elements and various types of single element behavior and interactions among multiple elements. Each result was 
completed between 
10-30 minutes, including the creation and testing time. They included different user interactions, such as following the mouse, using the arrow keys to control moving directions, using other keys to control moving speed, and using the mouse click to trigger dynamic effects. Multiple graphical controls {were used}, 
including user-drawn points for specifying target positions, lines for defining curves for moving paths, and regions for active effects. {For example,} the participants were able to define specific points and draw curves, simulating the movement of the sun along a designated path (Figure \ref{fig:result}(f)). They could also create defined areas, such as a {region} 
within a tree hollow where squirrels could move freely, with nuts appearing randomly in that region (Figure \ref{fig:result}(b)), enhancing the scene's liveliness. Additionally, users set up clickable regions (Figure \ref{fig:result}(g)), such as an orange that, when clicked, displayed an image of the fruit, fostering engagement and interaction. The results showed that users could create and edit various individual objects while facilitating interactions between them. They utilized text inputs to modify or redefine the behavior of single elements and employed automatically generated sliders to fine-tune specific details. Importantly, the operations on individual objects did not affect others or their interactions, ensuring clarity and control in complex scenarios. Overall, the findings indicate that our \sysName~system effectively supports user creativity and exploration in graphical interactions, making the creative process both enjoyable and intuitive.

\begin{figure*}[t]
\includegraphics[width=0.85\linewidth]{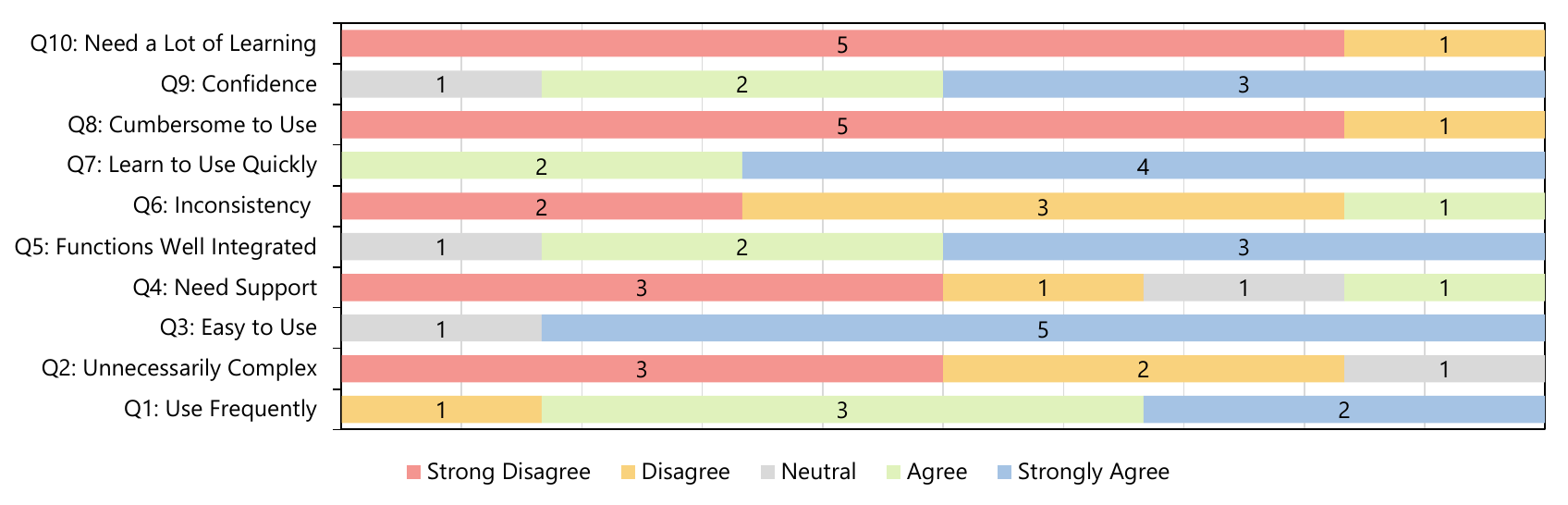}
  \caption{SUS score distribution. The question description is the key points from the full SUS questions.}
  \label{fig:sus}
\end{figure*}

The SUS score rated by the participants 
is 85 on a 100-point scale, indicating our system has good usability. The distribution of the SUS score for each question is shown in Figure \ref{fig:sus}. We observed that both individuals with programming backgrounds and those without found it easy to use our system to create their desired outcomes. The participants without programming experience (P2, P5-6) expressed a strong need for a tool that does not require coding since the cost of learning programming is quite high. Our system offers an intuitive solution, enabling them to design animations or games of varying complexity for use in their learning and daily life (P2, P6).
These users particularly noted that, beyond entertainment, our tool could also be employed to create diagrams and dynamic effects for academic papers (P3, P6). For instance, one participant with an environmental science background (P6) mentioned that generating dynamic visualizations for his papers was often challenging. With our system, he could quickly and easily create effective graphics to illustrate the core concepts of his work (Figure \ref{fig:result}(h)).

The participants with programming experience also found our system very useful. One participant with 8-year programming experience, P3, noted that if she were to create a game using traditional methods, it would take her at least two hours, but with our system, she completed it in just 15 minutes. This significantly saved {her} 
time, effort, and the need to learn new programming languages or frameworks, as she could directly generate game code through natural language and graphic controls. The participants (P1, P4) mentioned that our system did not require a technical person to guide them closely; a simple introduction and brief instructions at the start were sufficient for them to get up to speed quickly. 
{P1 particularly enjoyed that our system offers a ``modular architecture''. This capability allows users to navigate complex relationships among objects more effectively. P1 said, ``even for programming simple games, this system is much more easier to use than ChatGPT''. }

However, the participants also highlighted some limitations of our system, such as the currently limited graphic controls (P5-6), which only include points, straight lines, curves, and user-drawn regions. They desired 
additional geometric shapes, such as circles, rectangles, and triangles, as well as the ability to automatically segment imported background images. The ability of code generation is also questioned by participants (P2-3) in some aspects including adding unnecessary conditions. For example, in creating Scavenger hunt Game (Figure \ref{fig:result}(c)), when scripting the effect of a key appears and always follows Mario after Mario comes close to the box, it first generates the result that the key only follows Mario when Mario is close to the box. It adds a condition determination on whether Mario collides with the box, while it is unnecessary in the participant's expectation.

\section{Discussions and Future Work}

\subsection{Modularization with LLMs}
We tackled the challenge of interdependent effects on multiple elements within conversational LLMs through our \yh{element-level} modularization technique. While our primary focus is on creating 2D interactive scenes, this issue is prevalent across {many other} 
LLM applications. Despite advancements in LLM models, the problem of interdependence is likely to persist for the foreseeable future and is not confined to specific iterations, such as different versions of GPT. Our 
technique presents a novel and valuable approach applicable to {other} 
LLM-based frameworks {(e.g., Google Gemini, Claude, LLaMA)}, extending beyond just 2D visual creation to encompass a broader range of interactive scenarios.

\subsection{Trade-off between Modularization and Original ChatGPT}
Our modularization technique encourages users to provide descriptions for individual elements separately. While this promotes clarity and specificity, it can hinder seamless interaction to some extent. Users are required to manually delineate the behaviors of each element, in contrast to the original ChatGPT interface, where they can combine these elements more fluidly, although it often results in undesired and uncontrolled outcomes.

Looking ahead, we plan to develop an advanced input parsing and segmentation method that will allow users to input their descriptions without the need for strict separation. Our system will intelligently analyze and partition the descriptions of both individual and multiple elements, feeding them into distinct modules as necessary. This enhancement aims to strike a balance between user control and the fluidity of interaction, ultimately improving the overall user experience.

\subsection{Limitation of Context-awareness}
Although we provide context for code generation, when tasks become extremely complex and conversations lengthen, the issue of context loss for individual element creation persists. The inherent token limits of LLMs restrict the amount of information that can be processed within a single session. As context accumulates, it contributes to the text prompt, which can lead to truncated or incomplete responses, ultimately hindering the model's ability to generate coherent and contextually relevant outputs.

To address these challenges, future work should focus on developing effective context management strategies that prioritize and summarize essential information while discarding less relevant details. Implementing dynamic context windowing could allow the model to allocate more tokens to critical portions of the conversation while compressing less relevant exchanges. Additionally, exploring chunking mechanisms to break down complex tasks into smaller, manageable parts could help preserve context during longer interactions. By enhancing context-awareness in these ways, we aim to improve the robustness and coherence of LLM outputs, even in complex scenarios.

\yh{
\subsection{Potential Use Cases}
The open-ended study indicated that \sysName~can be used for creating dynamic and interactive scenarios for diverse purposes, such as interactive gaming experiences, engaging educational tools, visual communication in research, and UI design. Beyond these applications, we envision its use in other domains. For example, users can create their own medication reminders, journaling tools, and interactive mood trackers, by creating 2D representations of medicine, feelings, and mood and then crafting interactivity for them. Multiple family members can co-create interactive shared albums by importing photos to \sysName~and designing visualization patterns for memory recording and photo management. In addition, museums and exhibition centers can leverage \sysName~to enable visitors to express their feelings by creating interactive featured items, which can be analyzed for insights into visitor preferences. Furthermore, \sysName~can facilitates interactive storytelling, enabling children to create their own books and share their narratives in a fun and engaging way. 

\subsection{Integrating Code Representations and Element Visualizations}
Currently, we do not display the generated code in our interface.
However, we plan to gradually introduce code features as users become more comfortable with the basics. This will help them understand the underlying logic of their creations and provide a pathway to more advanced programming concepts. Besides, we are considering the development of educational modules aimed at teaching programming fundamentals. These components will be designed to prepare users for future coding tasks through engaging scene creation.

Incorporating visual chips into the UI design presents a significant opportunity to enhance user interaction. Recent generative AI tools \cite{wang2024promptcharm,li2024evaluating} effectively utilize chips to visually associate text prompts with corresponding visual elements, allowing users to quickly identify and swap components such as subjects and adjectives. While our current UI lists raw text, exploring a chip-based interface in future iterations could greatly improve user experience and accessibility. By visually representing elements like “Mario,” “Platform,” and “P2,” we can make it easier for users to understand and manipulate their inputs.

\subsection{Scalabiltiy for Complex Scenarios}
As more elements are added to a scene, the behavior and interactions among them can become complex. Users may need to view and manage these elements, their relationships, and interaction logic. To support this, we can enhance our system by integrating a graph representation \cite{yan2023xcreation}, where each node represents an element and each edge represents their relationships, with additional connected lines indicating the next event logic. This allows us to visualize both element behaviors and interactions as attributed objects \cite{xia2016object}, after the system generates corresponding codes. Users can then easily reuse these behaviors and interactions by dragging objects to other elements in the graph, as well as directly reconnecting or deleting elements and edges for quick management.

{In our current implementation,} 
\sysName~supports creating results containing multiple scenes (e.g., Figure \ref{fig:result}(d)) by switching in the central module to manage the visibility of created elements. However, more complex applications may contain more scenes. Our current implementation involves two levels: scene and scene elements. We might add one more level for scene management, which opens modules for scenes.
Users can manually create scenes and add elements, or we can extract scenes and elements from users' text input. 
We can visualize scenes as workflow UIs, connected by key condition variables, allowing users to manipulate scene order and relationships directly. This will open new opportunities for exploring conditional layered and nested LLMs in various applications.

}

\section{Conclusion}
This paper has introduced \sysName, a novel LLM-based system to simplify the creation of 2D interactive scenes without coding from natural language input and graphical control. We utilized content analysis on video tutorials about creating interactive scenes using ChatGPT and existing AI coding tools, and distilled several issues including the lack of independent generation and refinement, graphical control, and precise refinement. Based on these findings, we proposed a context-aware modularization technique that processes textual descriptions through individual LLM modules, with a central module coordinating interactions, allowing for independent refinement of each element. Our graphical user interface combined these modular LLMs with advanced graphical controls, enabling seamless code generation for 2D interactive scenes and direct integration of graphical information. We conducted a comparative study between \sysName~and \yh{Cursor Composer}, and found \sysName~significantly reduced the time, prompt trials, and prompt lengths and achieved better graphical and precise control in creating interactive scenes. Another open-ended usability study demonstrated that \sysName~allowed users to create various desired scenes easily without the need for coding, benefiting diverse application scenarios.



\bibliographystyle{ACM-Reference-Format}
\bibliography{Reference}


\begin{thebibliography}{79}


\ifx \showCODEN    \undefined \def \showCODEN     #1{\unskip}     \fi
\ifx \showDOI      \undefined \def \showDOI       #1{#1}\fi
\ifx \showISBNx    \undefined \def \showISBNx     #1{\unskip}     \fi
\ifx \showISBNxiii \undefined \def \showISBNxiii  #1{\unskip}     \fi
\ifx \showISSN     \undefined \def \showISSN      #1{\unskip}     \fi
\ifx \showLCCN     \undefined \def \showLCCN      #1{\unskip}     \fi
\ifx \shownote     \undefined \def \shownote      #1{#1}          \fi
\ifx \showarticletitle \undefined \def \showarticletitle #1{#1}   \fi
\ifx \showURL      \undefined \def \showURL       {\relax}        \fi
\providecommand\bibfield[2]{#2}
\providecommand\bibinfo[2]{#2}
\providecommand\natexlab[1]{#1}
\providecommand\showeprint[2][]{arXiv:#2}

\bibitem[flu({[n.\,d.]})]%
        {flutter}
 \bibinfo{year}{[n.\,d.]}\natexlab{}.
\newblock \bibinfo{title}{Flutter - Build apps for any screen}.
\newblock \bibinfo{howpublished}{\url{https://flutter.dev/}}.
\newblock
\newblock
\shownote{Accessed: 2024-11-26}.


\bibitem[pyt({[n.\,d.]})]%
        {python_playground}
 \bibinfo{year}{[n.\,d.]}\natexlab{}.
\newblock \bibinfo{title}{Python Playground - Online Python IDE}.
\newblock \bibinfo{howpublished}{\url{https://programiz.pro/ide/python}}.
\newblock
\newblock
\shownote{Accessed: 2024-11-26}.


\bibitem[Angert et~al\mbox{.}(2023)]%
        {angert2023spellburst}
\bibfield{author}{\bibinfo{person}{Tyler Angert}, \bibinfo{person}{Miroslav Suzara}, \bibinfo{person}{Jenny Han}, \bibinfo{person}{Christopher Pondoc}, {and} \bibinfo{person}{Hariharan Subramonyam}.} \bibinfo{year}{2023}\natexlab{}.
\newblock \showarticletitle{Spellburst: A node-based interface for exploratory creative coding with natural language prompts}. In \bibinfo{booktitle}{\emph{Proceedings of the 36th Annual ACM Symposium on User Interface Software and Technology}}. \bibinfo{pages}{1--22}.
\newblock


\bibitem[Anjum et~al\mbox{.}(2024)]%
        {anjum2024ink}
\bibfield{author}{\bibinfo{person}{Asad Anjum}, \bibinfo{person}{Yuting Li}, \bibinfo{person}{Noelle Law}, \bibinfo{person}{Megan Charity}, {and} \bibinfo{person}{Julian Togelius}.} \bibinfo{year}{2024}\natexlab{}.
\newblock \showarticletitle{The Ink Splotch Effect: A case study on ChatGPT as a co-creative game designer}. In \bibinfo{booktitle}{\emph{Proceedings of the 19th International Conference on the Foundations of Digital Games}}. \bibinfo{pages}{1--15}.
\newblock


\bibitem[Arawjo et~al\mbox{.}(2023)]%
        {arawjo2023chainforge}
\bibfield{author}{\bibinfo{person}{Ian Arawjo}, \bibinfo{person}{Priyan Vaithilingam}, \bibinfo{person}{Martin Wattenberg}, {and} \bibinfo{person}{Elena Glassman}.} \bibinfo{year}{2023}\natexlab{}.
\newblock \showarticletitle{ChainForge: An open-source visual programming environment for prompt engineering}. In \bibinfo{booktitle}{\emph{Adjunct Proceedings of the 36th Annual ACM Symposium on User Interface Software and Technology}}. \bibinfo{pages}{1--3}.
\newblock


\bibitem[Artifacts(2024)]%
        {claud_artifacts}
\bibfield{author}{\bibinfo{person}{Claude Artifacts}.} \bibinfo{year}{2024}\natexlab{}.
\newblock \bibinfo{title}{Claude Artifacts}.
\newblock
\newblock
\urldef\tempurl%
\url{https://madewithclaude.com/}
\showURL{%
\tempurl}
\newblock
\shownote{Accessed: 2024-12-10}.


\bibitem[Besta et~al\mbox{.}(2024)]%
        {besta2024graph}
\bibfield{author}{\bibinfo{person}{Maciej Besta}, \bibinfo{person}{Nils Blach}, \bibinfo{person}{Ales Kubicek}, \bibinfo{person}{Robert Gerstenberger}, \bibinfo{person}{Michal Podstawski}, \bibinfo{person}{Lukas Gianinazzi}, \bibinfo{person}{Joanna Gajda}, \bibinfo{person}{Tomasz Lehmann}, \bibinfo{person}{Hubert Niewiadomski}, \bibinfo{person}{Piotr Nyczyk}, {et~al\mbox{.}}} \bibinfo{year}{2024}\natexlab{}.
\newblock \showarticletitle{Graph of thoughts: Solving elaborate problems with large language models}. In \bibinfo{booktitle}{\emph{Proceedings of the AAAI Conference on Artificial Intelligence}}, Vol.~\bibinfo{volume}{38}. \bibinfo{pages}{17682--17690}.
\newblock


\bibitem[Bimbatti et~al\mbox{.}(2023)]%
        {bimbatti2023can}
\bibfield{author}{\bibinfo{person}{Giorgio Bimbatti}, \bibinfo{person}{Daniela Fogli}, \bibinfo{person}{Luigi Gargioni}, {et~al\mbox{.}}} \bibinfo{year}{2023}\natexlab{}.
\newblock \showarticletitle{Can ChatGPT support end-user development of robot programs?}. In \bibinfo{booktitle}{\emph{CEUR Workshop Proceedings}}, Vol.~\bibinfo{volume}{3408}.
\newblock


\bibitem[Brade et~al\mbox{.}(2023)]%
        {brade2023promptify}
\bibfield{author}{\bibinfo{person}{Stephen Brade}, \bibinfo{person}{Bryan Wang}, \bibinfo{person}{Mauricio Sousa}, \bibinfo{person}{Sageev Oore}, {and} \bibinfo{person}{Tovi Grossman}.} \bibinfo{year}{2023}\natexlab{}.
\newblock \showarticletitle{Promptify: Text-to-image generation through interactive prompt exploration with large language models}. In \bibinfo{booktitle}{\emph{Proceedings of the 36th Annual ACM Symposium on User Interface Software and Technology}}. \bibinfo{pages}{1--14}.
\newblock


\bibitem[Charmaz(2008)]%
        {charmaz2008constructionism}
\bibfield{author}{\bibinfo{person}{Kathy Charmaz}.} \bibinfo{year}{2008}\natexlab{}.
\newblock \showarticletitle{Constructionism and the grounded theory method}.
\newblock \bibinfo{journal}{\emph{Handbook of constructionist research}} \bibinfo{volume}{1}, \bibinfo{number}{1} (\bibinfo{year}{2008}), \bibinfo{pages}{397--412}.
\newblock


\bibitem[Chen et~al\mbox{.}(2024)]%
        {chen2024chatscratch}
\bibfield{author}{\bibinfo{person}{Liuqing Chen}, \bibinfo{person}{Shuhong Xiao}, \bibinfo{person}{Yunnong Chen}, \bibinfo{person}{Yaxuan Song}, \bibinfo{person}{Ruoyu Wu}, {and} \bibinfo{person}{Lingyun Sun}.} \bibinfo{year}{2024}\natexlab{}.
\newblock \showarticletitle{ChatScratch: An AI-Augmented System Toward Autonomous Visual Programming Learning for Children Aged 6-12}. In \bibinfo{booktitle}{\emph{Proceedings of the CHI Conference on Human Factors in Computing Systems}}. \bibinfo{pages}{1--19}.
\newblock


\bibitem[Chen et~al\mbox{.}(2021)]%
        {chen2021entanglevr}
\bibfield{author}{\bibinfo{person}{Mengyu Chen}, \bibinfo{person}{Marko Peljhan}, {and} \bibinfo{person}{Misha Sra}.} \bibinfo{year}{2021}\natexlab{}.
\newblock \showarticletitle{Entanglevr: A visual programming interface for virtual reality interactive scene generation}. In \bibinfo{booktitle}{\emph{Proceedings of the 27th ACM symposium on virtual reality software and technology}}. \bibinfo{pages}{1--6}.
\newblock


\bibitem[{CodePen}(2012)]%
        {codepen}
\bibfield{author}{\bibinfo{person}{{CodePen}}.} \bibinfo{year}{2012}\natexlab{}.
\newblock \bibinfo{title}{CodePen}.
\newblock
\newblock
\urldef\tempurl%
\url{https://codepen.io}
\showURL{%
\tempurl}
\newblock
\shownote{Accessed: 2024-09-11}.


\bibitem[{Cursor}(2023)]%
        {cursor2023}
\bibfield{author}{\bibinfo{person}{{Cursor}}.} \bibinfo{year}{2023}\natexlab{}.
\newblock \bibinfo{title}{Cursor - The AI Code Editor}.
\newblock
\newblock
\urldef\tempurl%
\url{https://www.cursor.com/}
\showURL{%
\tempurl}
\newblock
\shownote{Accessed: 2024-11-30}.


\bibitem[Dang et~al\mbox{.}(2022)]%
        {dang2022prompt}
\bibfield{author}{\bibinfo{person}{Hai Dang}, \bibinfo{person}{Lukas Mecke}, \bibinfo{person}{Florian Lehmann}, \bibinfo{person}{Sven Goller}, {and} \bibinfo{person}{Daniel Buschek}.} \bibinfo{year}{2022}\natexlab{}.
\newblock \showarticletitle{How to prompt? Opportunities and challenges of zero-and few-shot learning for human-AI interaction in creative applications of generative models}.
\newblock \bibinfo{journal}{\emph{arXiv preprint arXiv:2209.01390}} (\bibinfo{year}{2022}).
\newblock


\bibitem[De~La~Torre et~al\mbox{.}(2024)]%
        {de2024llmr}
\bibfield{author}{\bibinfo{person}{Fernanda De~La~Torre}, \bibinfo{person}{Cathy~Mengying Fang}, \bibinfo{person}{Han Huang}, \bibinfo{person}{Andrzej Banburski-Fahey}, \bibinfo{person}{Judith Amores~Fernandez}, {and} \bibinfo{person}{Jaron Lanier}.} \bibinfo{year}{2024}\natexlab{}.
\newblock \showarticletitle{Llmr: Real-time prompting of interactive worlds using large language models}. In \bibinfo{booktitle}{\emph{Proceedings of the CHI Conference on Human Factors in Computing Systems}}. \bibinfo{pages}{1--22}.
\newblock


\bibitem[{Exploratorium}(1993)]%
        {exploratorium}
\bibfield{author}{\bibinfo{person}{{Exploratorium}}.} \bibinfo{year}{1993}\natexlab{}.
\newblock \bibinfo{title}{Exploratorium}.
\newblock
\newblock
\urldef\tempurl%
\url{https://www.exploratorium.edu}
\showURL{%
\tempurl}
\newblock
\shownote{Accessed: 2024-09-11}.


\bibitem[{GitHub}(2021)]%
        {github_copilot}
\bibfield{author}{\bibinfo{person}{{GitHub}}.} \bibinfo{year}{2021}\natexlab{}.
\newblock \bibinfo{title}{GitHub Copilot}.
\newblock
\newblock
\urldef\tempurl%
\url{https://github.com/features/copilot}
\showURL{%
\tempurl}
\newblock
\shownote{Accessed: 2024-09-11}.


\bibitem[{Glitch Team}(2017)]%
        {glitch}
\bibfield{author}{\bibinfo{person}{{Glitch Team}}.} \bibinfo{year}{2017}\natexlab{}.
\newblock \bibinfo{title}{Glitch}.
\newblock
\newblock
\urldef\tempurl%
\url{https://glitch.com}
\showURL{%
\tempurl}
\newblock
\shownote{Accessed: 2024-09-11}.


\bibitem[Harwood and Garry(2003)]%
        {harwood2003overview}
\bibfield{author}{\bibinfo{person}{Tracy~G Harwood} {and} \bibinfo{person}{Tony Garry}.} \bibinfo{year}{2003}\natexlab{}.
\newblock \showarticletitle{An overview of content analysis}.
\newblock \bibinfo{journal}{\emph{The marketing review}} \bibinfo{volume}{3}, \bibinfo{number}{4} (\bibinfo{year}{2003}), \bibinfo{pages}{479--498}.
\newblock


\bibitem[Hou et~al\mbox{.}(2024)]%
        {hou2024c2ideas}
\bibfield{author}{\bibinfo{person}{Yihan Hou}, \bibinfo{person}{Manling Yang}, \bibinfo{person}{Hao Cui}, \bibinfo{person}{Lei Wang}, \bibinfo{person}{Jie Xu}, {and} \bibinfo{person}{Wei Zeng}.} \bibinfo{year}{2024}\natexlab{}.
\newblock \showarticletitle{C2Ideas: Supporting Creative Interior Color Design Ideation with a Large Language Model}. In \bibinfo{booktitle}{\emph{Proceedings of the CHI Conference on Human Factors in Computing Systems}}. \bibinfo{pages}{1--18}.
\newblock


\bibitem[Hu et~al\mbox{.}(2024)]%
        {hu2024generating}
\bibfield{author}{\bibinfo{person}{Chengpeng Hu}, \bibinfo{person}{Yunlong Zhao}, {and} \bibinfo{person}{Jialin Liu}.} \bibinfo{year}{2024}\natexlab{}.
\newblock \showarticletitle{Generating Games via LLMs: An Investigation with Video Game Description Language}.
\newblock \bibinfo{journal}{\emph{arXiv preprint arXiv:2404.08706}} (\bibinfo{year}{2024}).
\newblock


\bibitem[Huang et~al\mbox{.}(2024)]%
        {huang2024anpl}
\bibfield{author}{\bibinfo{person}{Di Huang}, \bibinfo{person}{Ziyuan Nan}, \bibinfo{person}{Xing Hu}, \bibinfo{person}{Pengwei Jin}, \bibinfo{person}{Shaohui Peng}, \bibinfo{person}{Yuanbo Wen}, \bibinfo{person}{Rui Zhang}, \bibinfo{person}{Zidong Du}, \bibinfo{person}{Qi Guo}, \bibinfo{person}{Yewen Pu}, {et~al\mbox{.}}} \bibinfo{year}{2024}\natexlab{}.
\newblock \showarticletitle{ANPL: towards natural programming with interactive decomposition}.
\newblock \bibinfo{journal}{\emph{Advances in Neural Information Processing Systems}}  \bibinfo{volume}{36} (\bibinfo{year}{2024}).
\newblock


\bibitem[Jiang et~al\mbox{.}(2023)]%
        {jiang2023graphologue}
\bibfield{author}{\bibinfo{person}{Peiling Jiang}, \bibinfo{person}{Jude Rayan}, \bibinfo{person}{Steven~P Dow}, {and} \bibinfo{person}{Haijun Xia}.} \bibinfo{year}{2023}\natexlab{}.
\newblock \showarticletitle{Graphologue: Exploring large language model responses with interactive diagrams}. In \bibinfo{booktitle}{\emph{Proceedings of the 36th Annual ACM Symposium on User Interface Software and Technology}}. \bibinfo{pages}{1--20}.
\newblock


\bibitem[Kazi et~al\mbox{.}(2014a)]%
        {kazi2014kitty}
\bibfield{author}{\bibinfo{person}{Rubaiat~Habib Kazi}, \bibinfo{person}{Fanny Chevalier}, \bibinfo{person}{Tovi Grossman}, {and} \bibinfo{person}{George Fitzmaurice}.} \bibinfo{year}{2014}\natexlab{a}.
\newblock \showarticletitle{Kitty: sketching dynamic and interactive illustrations}. In \bibinfo{booktitle}{\emph{Proceedings of the 27th annual ACM symposium on User interface software and technology}}. \bibinfo{pages}{395--405}.
\newblock


\bibitem[Kazi et~al\mbox{.}(2014b)]%
        {kazi2014draco}
\bibfield{author}{\bibinfo{person}{Rubaiat~Habib Kazi}, \bibinfo{person}{Fanny Chevalier}, \bibinfo{person}{Tovi Grossman}, \bibinfo{person}{Shengdong Zhao}, {and} \bibinfo{person}{George Fitzmaurice}.} \bibinfo{year}{2014}\natexlab{b}.
\newblock \showarticletitle{Draco: bringing life to illustrations with kinetic textures}. In \bibinfo{booktitle}{\emph{Proceedings of the SIGCHI Conference on Human Factors in Computing Systems}}. \bibinfo{pages}{351--360}.
\newblock


\bibitem[{Khan Academy}(2008)]%
        {khanacademy}
\bibfield{author}{\bibinfo{person}{{Khan Academy}}.} \bibinfo{year}{2008}\natexlab{}.
\newblock \bibinfo{title}{Khan Academy}.
\newblock
\newblock
\urldef\tempurl%
\url{https://www.khanacademy.org}
\showURL{%
\tempurl}
\newblock
\shownote{Accessed: 2024-09-11}.


\bibitem[Kim et~al\mbox{.}(2023)]%
        {kim2023cells}
\bibfield{author}{\bibinfo{person}{Tae~Soo Kim}, \bibinfo{person}{Yoonjoo Lee}, \bibinfo{person}{Minsuk Chang}, {and} \bibinfo{person}{Juho Kim}.} \bibinfo{year}{2023}\natexlab{}.
\newblock \showarticletitle{Cells, generators, and lenses: Design framework for object-oriented interaction with large language models}. In \bibinfo{booktitle}{\emph{Proceedings of the 36th Annual ACM Symposium on User Interface Software and Technology}}. \bibinfo{pages}{1--18}.
\newblock


\bibitem[Lan et~al\mbox{.}(2023)]%
        {lan2023application}
\bibfield{author}{\bibinfo{person}{Chong Lan}, \bibinfo{person}{Yongsheng Wang}, \bibinfo{person}{Chengze Wang}, \bibinfo{person}{Shirong Song}, {and} \bibinfo{person}{Zheng Gong}.} \bibinfo{year}{2023}\natexlab{}.
\newblock \showarticletitle{Application of ChatGPT-Based Digital Human in Animation Creation}.
\newblock \bibinfo{journal}{\emph{Future Internet}} \bibinfo{volume}{15}, \bibinfo{number}{9} (\bibinfo{year}{2023}), \bibinfo{pages}{300}.
\newblock


\bibitem[Li et~al\mbox{.}(2024)]%
        {li2024evaluating}
\bibfield{author}{\bibinfo{person}{Baiqi Li}, \bibinfo{person}{Zhiqiu Lin}, \bibinfo{person}{Deepak Pathak}, \bibinfo{person}{Jiayao Li}, \bibinfo{person}{Yixin Fei}, \bibinfo{person}{Kewen Wu}, \bibinfo{person}{Xide Xia}, \bibinfo{person}{Pengchuan Zhang}, \bibinfo{person}{Graham Neubig}, {and} \bibinfo{person}{Deva Ramanan}.} \bibinfo{year}{2024}\natexlab{}.
\newblock \showarticletitle{Evaluating and Improving Compositional Text-to-Visual Generation}. In \bibinfo{booktitle}{\emph{Proceedings of the IEEE/CVF Conference on Computer Vision and Pattern Recognition}}. \bibinfo{pages}{5290--5301}.
\newblock


\bibitem[Li et~al\mbox{.}(2023)]%
        {li2023chatiot}
\bibfield{author}{\bibinfo{person}{Fu Li}, \bibinfo{person}{Jiaming Huang}, \bibinfo{person}{Yi Gao}, {and} \bibinfo{person}{Wei Dong}.} \bibinfo{year}{2023}\natexlab{}.
\newblock \showarticletitle{ChatIoT: Zero-code Generation of Trigger-action Based IoT Programs with ChatGPT}. In \bibinfo{booktitle}{\emph{Proceedings of the 7th Asia-Pacific Workshop on Networking}}. \bibinfo{pages}{219--220}.
\newblock


\bibitem[Li et~al\mbox{.}(2022)]%
        {li2022competition}
\bibfield{author}{\bibinfo{person}{Yujia Li}, \bibinfo{person}{David Choi}, \bibinfo{person}{Junyoung Chung}, \bibinfo{person}{Nate Kushman}, \bibinfo{person}{Julian Schrittwieser}, \bibinfo{person}{R{\'e}mi Leblond}, \bibinfo{person}{Tom Eccles}, \bibinfo{person}{James Keeling}, \bibinfo{person}{Felix Gimeno}, \bibinfo{person}{Agustin Dal~Lago}, {et~al\mbox{.}}} \bibinfo{year}{2022}\natexlab{}.
\newblock \showarticletitle{Competition-level code generation with alphacode}.
\newblock \bibinfo{journal}{\emph{Science}} \bibinfo{volume}{378}, \bibinfo{number}{6624} (\bibinfo{year}{2022}), \bibinfo{pages}{1092--1097}.
\newblock


\bibitem[Liu et~al\mbox{.}(2020)]%
        {liu2020posetween}
\bibfield{author}{\bibinfo{person}{Jingyuan Liu}, \bibinfo{person}{Hongbo Fu}, {and} \bibinfo{person}{Chiew-Lan Tai}.} \bibinfo{year}{2020}\natexlab{}.
\newblock \showarticletitle{Posetween: Pose-driven tween animation}. In \bibinfo{booktitle}{\emph{Proceedings of the 33rd annual acm symposium on user interface software and technology}}. \bibinfo{pages}{791--804}.
\newblock


\bibitem[Liu et~al\mbox{.}(2024a)]%
        {liu2024logomotion}
\bibfield{author}{\bibinfo{person}{Vivian Liu}, \bibinfo{person}{Rubaiat~Habib Kazi}, \bibinfo{person}{Li-Yi Wei}, \bibinfo{person}{Matthew Fisher}, \bibinfo{person}{Timothy Langlois}, \bibinfo{person}{Seth Walker}, {and} \bibinfo{person}{Lydia Chilton}.} \bibinfo{year}{2024}\natexlab{a}.
\newblock \showarticletitle{LogoMotion: Visually Grounded Code Generation for Content-Aware Animation}.
\newblock \bibinfo{journal}{\emph{arXiv preprint arXiv:2405.07065}} (\bibinfo{year}{2024}).
\newblock


\bibitem[Liu et~al\mbox{.}(2022)]%
        {liu2022opal}
\bibfield{author}{\bibinfo{person}{Vivian Liu}, \bibinfo{person}{Han Qiao}, {and} \bibinfo{person}{Lydia Chilton}.} \bibinfo{year}{2022}\natexlab{}.
\newblock \showarticletitle{Opal: Multimodal image generation for news illustration}. In \bibinfo{booktitle}{\emph{Proceedings of the 35th Annual ACM Symposium on User Interface Software and Technology}}. \bibinfo{pages}{1--17}.
\newblock


\bibitem[Liu et~al\mbox{.}(2024b)]%
        {liu2024refining}
\bibfield{author}{\bibinfo{person}{Yue Liu}, \bibinfo{person}{Thanh Le-Cong}, \bibinfo{person}{Ratnadira Widyasari}, \bibinfo{person}{Chakkrit Tantithamthavorn}, \bibinfo{person}{Li Li}, \bibinfo{person}{Xuan-Bach~D Le}, {and} \bibinfo{person}{David Lo}.} \bibinfo{year}{2024}\natexlab{b}.
\newblock \showarticletitle{Refining chatgpt-generated code: Characterizing and mitigating code quality issues}.
\newblock \bibinfo{journal}{\emph{ACM Transactions on Software Engineering and Methodology}} \bibinfo{volume}{33}, \bibinfo{number}{5} (\bibinfo{year}{2024}), \bibinfo{pages}{1--26}.
\newblock


\bibitem[Liu et~al\mbox{.}(2024c)]%
        {liu2024no}
\bibfield{author}{\bibinfo{person}{Zhijie Liu}, \bibinfo{person}{Yutian Tang}, \bibinfo{person}{Xiapu Luo}, \bibinfo{person}{Yuming Zhou}, {and} \bibinfo{person}{Liang~Feng Zhang}.} \bibinfo{year}{2024}\natexlab{c}.
\newblock \showarticletitle{No need to lift a finger anymore? assessing the quality of code generation by chatgpt}.
\newblock \bibinfo{journal}{\emph{IEEE Transactions on Software Engineering}} (\bibinfo{year}{2024}).
\newblock


\bibitem[Masson et~al\mbox{.}(2024)]%
        {masson2024directgpt}
\bibfield{author}{\bibinfo{person}{Damien Masson}, \bibinfo{person}{Sylvain Malacria}, \bibinfo{person}{G{\'e}ry Casiez}, {and} \bibinfo{person}{Daniel Vogel}.} \bibinfo{year}{2024}\natexlab{}.
\newblock \showarticletitle{Directgpt: A direct manipulation interface to interact with large language models}. In \bibinfo{booktitle}{\emph{Proceedings of the CHI Conference on Human Factors in Computing Systems}}. \bibinfo{pages}{1--16}.
\newblock


\bibitem[{MIT Media Lab}(2003)]%
        {scratch}
\bibfield{author}{\bibinfo{person}{{MIT Media Lab}}.} \bibinfo{year}{2003}\natexlab{}.
\newblock \bibinfo{title}{Scratch}.
\newblock
\newblock
\urldef\tempurl%
\url{https://scratch.mit.edu}
\showURL{%
\tempurl}
\newblock
\shownote{Accessed: 2024-09-11}.


\bibitem[Myers et~al\mbox{.}(2008)]%
        {myers2008designers}
\bibfield{author}{\bibinfo{person}{Brad Myers}, \bibinfo{person}{Sun~Young Park}, \bibinfo{person}{Yoko Nakano}, \bibinfo{person}{Greg Mueller}, {and} \bibinfo{person}{Amy Ko}.} \bibinfo{year}{2008}\natexlab{}.
\newblock \showarticletitle{How designers design and program interactive behaviors}. In \bibinfo{booktitle}{\emph{2008 IEEE Symposium on Visual Languages and Human-Centric Computing}}. IEEE, \bibinfo{pages}{177--184}.
\newblock


\bibitem[Nijkamp et~al\mbox{.}(2022)]%
        {nijkamp2022codegen}
\bibfield{author}{\bibinfo{person}{Erik Nijkamp}, \bibinfo{person}{Bo Pang}, \bibinfo{person}{Hiroaki Hayashi}, \bibinfo{person}{Lifu Tu}, \bibinfo{person}{Huan Wang}, \bibinfo{person}{Yingbo Zhou}, \bibinfo{person}{Silvio Savarese}, {and} \bibinfo{person}{Caiming Xiong}.} \bibinfo{year}{2022}\natexlab{}.
\newblock \showarticletitle{Codegen: An open large language model for code with multi-turn program synthesis}.
\newblock \bibinfo{journal}{\emph{arXiv preprint arXiv:2203.13474}} (\bibinfo{year}{2022}).
\newblock


\bibitem[{Nintendo}(1986)]%
        {zelda}
\bibfield{author}{\bibinfo{person}{{Nintendo}}.} \bibinfo{year}{1986}\natexlab{}.
\newblock \bibinfo{title}{The Legend of Zelda}.
\newblock
\newblock
\urldef\tempurl%
\url{https://en.wikipedia.org/wiki/The_Legend_of_Zelda}
\showURL{%
\tempurl}
\newblock
\shownote{Accessed: 2024-09-11}.


\bibitem[{OpenAI}(2023)]%
        {openai_chatgpt}
\bibfield{author}{\bibinfo{person}{{OpenAI}}.} \bibinfo{year}{2023}\natexlab{}.
\newblock \bibinfo{title}{ChatGPT}.
\newblock
\newblock
\urldef\tempurl%
\url{https://chat.openai.com}
\showURL{%
\tempurl}
\newblock
\shownote{Accessed: 2024-09-11}.


\bibitem[{Pajitnov, Alexey}(1984)]%
        {tetris}
\bibfield{author}{\bibinfo{person}{{Pajitnov, Alexey}}.} \bibinfo{year}{1984}\natexlab{}.
\newblock \bibinfo{title}{Tetris}.
\newblock
\newblock
\urldef\tempurl%
\url{https://en.wikipedia.org/wiki/Tetris}
\showURL{%
\tempurl}
\newblock
\shownote{Accessed: 2024-09-11}.


\bibitem[{Phaser Team}(2013)]%
        {phaser}
\bibfield{author}{\bibinfo{person}{{Phaser Team}}.} \bibinfo{year}{2013}\natexlab{}.
\newblock \bibinfo{title}{Phaser}.
\newblock
\newblock
\urldef\tempurl%
\url{https://phaser.io}
\showURL{%
\tempurl}
\newblock
\shownote{Accessed: 2024-09-11}.


\bibitem[{Project Jupyter}(2014)]%
        {jupyter}
\bibfield{author}{\bibinfo{person}{{Project Jupyter}}.} \bibinfo{year}{2014}\natexlab{}.
\newblock \bibinfo{title}{Jupyter Notebook}.
\newblock
\newblock
\urldef\tempurl%
\url{https://jupyter.org}
\showURL{%
\tempurl}
\newblock
\shownote{Accessed: 2024-09-11}.


\bibitem[Qian et~al\mbox{.}(2024)]%
        {qian2024chatdev}
\bibfield{author}{\bibinfo{person}{Chen Qian}, \bibinfo{person}{Wei Liu}, \bibinfo{person}{Hongzhang Liu}, \bibinfo{person}{Nuo Chen}, \bibinfo{person}{Yufan Dang}, \bibinfo{person}{Jiahao Li}, \bibinfo{person}{Cheng Yang}, \bibinfo{person}{Weize Chen}, \bibinfo{person}{Yusheng Su}, \bibinfo{person}{Xin Cong}, {et~al\mbox{.}}} \bibinfo{year}{2024}\natexlab{}.
\newblock \showarticletitle{Chatdev: Communicative agents for software development}. In \bibinfo{booktitle}{\emph{Proceedings of the 62nd Annual Meeting of the Association for Computational Linguistics (Volume 1: Long Papers)}}. \bibinfo{pages}{15174--15186}.
\newblock


\bibitem[Ritschel et~al\mbox{.}(2022)]%
        {ritschel2022can}
\bibfield{author}{\bibinfo{person}{Nico Ritschel}, \bibinfo{person}{Felipe Fronchetti}, \bibinfo{person}{Reid Holmes}, \bibinfo{person}{Ronald Garcia}, {and} \bibinfo{person}{David~C Shepherd}.} \bibinfo{year}{2022}\natexlab{}.
\newblock \showarticletitle{Can guided decomposition help end-users write larger block-based programs? a mobile robot experiment}.
\newblock \bibinfo{journal}{\emph{Proceedings of the ACM on Programming Languages}} \bibinfo{volume}{6}, \bibinfo{number}{OOPSLA2} (\bibinfo{year}{2022}), \bibinfo{pages}{233--258}.
\newblock


\bibitem[Rosenberg et~al\mbox{.}(2024)]%
        {rosenberg2024drawtalking}
\bibfield{author}{\bibinfo{person}{Karl~Toby Rosenberg}, \bibinfo{person}{Rubaiat~Habib Kazi}, \bibinfo{person}{Li-Yi Wei}, \bibinfo{person}{Haijun Xia}, {and} \bibinfo{person}{Ken Perlin}.} \bibinfo{year}{2024}\natexlab{}.
\newblock \showarticletitle{DrawTalking: Building Interactive Worlds by Sketching and Speaking}.
\newblock \bibinfo{journal}{\emph{arXiv preprint arXiv:2401.05631}} (\bibinfo{year}{2024}).
\newblock


\bibitem[Roziere et~al\mbox{.}(2023)]%
        {roziere2023code}
\bibfield{author}{\bibinfo{person}{Baptiste Roziere}, \bibinfo{person}{Jonas Gehring}, \bibinfo{person}{Fabian Gloeckle}, \bibinfo{person}{Sten Sootla}, \bibinfo{person}{Itai Gat}, \bibinfo{person}{Xiaoqing~Ellen Tan}, \bibinfo{person}{Yossi Adi}, \bibinfo{person}{Jingyu Liu}, \bibinfo{person}{Tal Remez}, \bibinfo{person}{J{\'e}r{\'e}my Rapin}, {et~al\mbox{.}}} \bibinfo{year}{2023}\natexlab{}.
\newblock \showarticletitle{Code llama: Open foundation models for code}.
\newblock \bibinfo{journal}{\emph{arXiv preprint arXiv:2308.12950}} (\bibinfo{year}{2023}).
\newblock


\bibitem[Siddiq et~al\mbox{.}(2024)]%
        {siddiq2024quality}
\bibfield{author}{\bibinfo{person}{Mohammed~Latif Siddiq}, \bibinfo{person}{Lindsay Roney}, \bibinfo{person}{Jiahao Zhang}, {and} \bibinfo{person}{Joanna Cecilia Da~Silva Santos}.} \bibinfo{year}{2024}\natexlab{}.
\newblock \showarticletitle{Quality Assessment of ChatGPT Generated Code and their Use by Developers}. In \bibinfo{booktitle}{\emph{Proceedings of the 21st International Conference on Mining Software Repositories}}. \bibinfo{pages}{152--156}.
\newblock


\bibitem[{Snap}(2023)]%
        {snap_lensstudio}
\bibfield{author}{\bibinfo{person}{{Snap}}.} \bibinfo{year}{2023}\natexlab{}.
\newblock \bibinfo{title}{Lens Studio}.
\newblock
\newblock
\urldef\tempurl%
\url{https://ar.snap.com/en-US/lens-studio}
\showURL{%
\tempurl}
\newblock
\shownote{Accessed: 2024-09-11}.


\bibitem[Sudhakaran et~al\mbox{.}(2024)]%
        {sudhakaran2024mariogpt}
\bibfield{author}{\bibinfo{person}{Shyam Sudhakaran}, \bibinfo{person}{Miguel Gonz{\'a}lez-Duque}, \bibinfo{person}{Matthias Freiberger}, \bibinfo{person}{Claire Glanois}, \bibinfo{person}{Elias Najarro}, {and} \bibinfo{person}{Sebastian Risi}.} \bibinfo{year}{2024}\natexlab{}.
\newblock \showarticletitle{Mariogpt: Open-ended text2level generation through large language models}.
\newblock \bibinfo{journal}{\emph{Advances in Neural Information Processing Systems}}  \bibinfo{volume}{36} (\bibinfo{year}{2024}).
\newblock


\bibitem[Sweetser(2024)]%
        {sweetser2024large}
\bibfield{author}{\bibinfo{person}{Penny Sweetser}.} \bibinfo{year}{2024}\natexlab{}.
\newblock \showarticletitle{Large language models and video games: A preliminary scoping review}. In \bibinfo{booktitle}{\emph{Proceedings of the 6th ACM Conference on Conversational User Interfaces}}. \bibinfo{pages}{1--8}.
\newblock


\bibitem[Tian et~al\mbox{.}(2023)]%
        {tian2023chatgpt}
\bibfield{author}{\bibinfo{person}{Haoye Tian}, \bibinfo{person}{Weiqi Lu}, \bibinfo{person}{Tsz~On Li}, \bibinfo{person}{Xunzhu Tang}, \bibinfo{person}{Shing-Chi Cheung}, \bibinfo{person}{Jacques Klein}, {and} \bibinfo{person}{Tegawend{\'e}~F Bissyand{\'e}}.} \bibinfo{year}{2023}\natexlab{}.
\newblock \showarticletitle{Is ChatGPT the ultimate programming assistant--how far is it?}
\newblock \bibinfo{journal}{\emph{arXiv preprint arXiv:2304.11938}} (\bibinfo{year}{2023}).
\newblock


\bibitem[Todd et~al\mbox{.}(2023)]%
        {todd2023level}
\bibfield{author}{\bibinfo{person}{Graham Todd}, \bibinfo{person}{Sam Earle}, \bibinfo{person}{Muhammad~Umair Nasir}, \bibinfo{person}{Michael~Cerny Green}, {and} \bibinfo{person}{Julian Togelius}.} \bibinfo{year}{2023}\natexlab{}.
\newblock \showarticletitle{Level generation through large language models}. In \bibinfo{booktitle}{\emph{Proceedings of the 18th International Conference on the Foundations of Digital Games}}. \bibinfo{pages}{1--8}.
\newblock


\bibitem[Tseng et~al\mbox{.}(2024)]%
        {tseng2024keyframer}
\bibfield{author}{\bibinfo{person}{Tiffany Tseng}, \bibinfo{person}{Ruijia Cheng}, {and} \bibinfo{person}{Jeffrey Nichols}.} \bibinfo{year}{2024}\natexlab{}.
\newblock \showarticletitle{Keyframer: Empowering Animation Design using Large Language Models}.
\newblock \bibinfo{journal}{\emph{arXiv preprint arXiv:2402.06071}} (\bibinfo{year}{2024}).
\newblock


\bibitem[{Unity Technologies}(2005)]%
        {unity}
\bibfield{author}{\bibinfo{person}{{Unity Technologies}}.} \bibinfo{year}{2005}\natexlab{}.
\newblock \bibinfo{title}{Unity}.
\newblock
\newblock
\urldef\tempurl%
\url{https://unity.com}
\showURL{%
\tempurl}
\newblock
\shownote{Accessed: 2024-09-11}.


\bibitem[{University of Colorado Boulder}(2002)]%
        {phET}
\bibfield{author}{\bibinfo{person}{{University of Colorado Boulder}}.} \bibinfo{year}{2002}\natexlab{}.
\newblock \bibinfo{title}{PhET Interactive Simulations}.
\newblock
\newblock
\urldef\tempurl%
\url{https://phet.colorado.edu}
\showURL{%
\tempurl}
\newblock
\shownote{Accessed: 2024-09-11}.


\bibitem[{Unreal Engine}(2023)]%
        {unreal_blueprints}
\bibfield{author}{\bibinfo{person}{{Unreal Engine}}.} \bibinfo{year}{2023}\natexlab{}.
\newblock \bibinfo{title}{Introduction to Blueprints}.
\newblock
\newblock
\urldef\tempurl%
\url{https://docs.unrealengine.com/4.27/en-US/ProgrammingAndScripting/Blueprints/GettingStarted/}
\showURL{%
\tempurl}
\newblock
\shownote{Accessed: 2024-09-11}.


\bibitem[Vercel(2024)]%
        {v0_vercel}
\bibfield{author}{\bibinfo{person}{Vercel}.} \bibinfo{year}{2024}\natexlab{}.
\newblock \bibinfo{title}{v0 by Vercel}.
\newblock
\newblock
\urldef\tempurl%
\url{https://v0.dev/}
\showURL{%
\tempurl}
\newblock
\shownote{Accessed: 2024-12-10}.


\bibitem[Wang et~al\mbox{.}(2024)]%
        {wang2024promptcharm}
\bibfield{author}{\bibinfo{person}{Zhijie Wang}, \bibinfo{person}{Yuheng Huang}, \bibinfo{person}{Da Song}, \bibinfo{person}{Lei Ma}, {and} \bibinfo{person}{Tianyi Zhang}.} \bibinfo{year}{2024}\natexlab{}.
\newblock \showarticletitle{PromptCharm: Text-to-Image Generation through Multi-modal Prompting and Refinement}. In \bibinfo{booktitle}{\emph{Proceedings of the CHI Conference on Human Factors in Computing Systems}}. \bibinfo{pages}{1--21}.
\newblock


\bibitem[Wei et~al\mbox{.}(2022)]%
        {wei2022chain}
\bibfield{author}{\bibinfo{person}{Jason Wei}, \bibinfo{person}{Xuezhi Wang}, \bibinfo{person}{Dale Schuurmans}, \bibinfo{person}{Maarten Bosma}, \bibinfo{person}{Fei Xia}, \bibinfo{person}{Ed Chi}, \bibinfo{person}{Quoc~V Le}, \bibinfo{person}{Denny Zhou}, {et~al\mbox{.}}} \bibinfo{year}{2022}\natexlab{}.
\newblock \showarticletitle{Chain-of-thought prompting elicits reasoning in large language models}.
\newblock \bibinfo{journal}{\emph{Advances in neural information processing systems}}  \bibinfo{volume}{35} (\bibinfo{year}{2022}), \bibinfo{pages}{24824--24837}.
\newblock


\bibitem[{Wikipedia contributors}(2023)]%
        {supermario}
\bibfield{author}{\bibinfo{person}{{Wikipedia contributors}}.} \bibinfo{year}{2023}\natexlab{}.
\newblock \bibinfo{title}{Super Mario}.
\newblock
\newblock
\urldef\tempurl%
\url{https://en.wikipedia.org/wiki/Super_Mario}
\showURL{%
\tempurl}
\newblock
\shownote{Accessed: 2024-09-11}.


\bibitem[Wu et~al\mbox{.}(2023)]%
        {wu2023visual}
\bibfield{author}{\bibinfo{person}{Chenfei Wu}, \bibinfo{person}{Shengming Yin}, \bibinfo{person}{Weizhen Qi}, \bibinfo{person}{Xiaodong Wang}, \bibinfo{person}{Zecheng Tang}, {and} \bibinfo{person}{Nan Duan}.} \bibinfo{year}{2023}\natexlab{}.
\newblock \showarticletitle{Visual chatgpt: Talking, drawing and editing with visual foundation models}.
\newblock \bibinfo{journal}{\emph{arXiv preprint arXiv:2303.04671}} (\bibinfo{year}{2023}).
\newblock


\bibitem[Wu et~al\mbox{.}(2022a)]%
        {wu2022promptchainer}
\bibfield{author}{\bibinfo{person}{Tongshuang Wu}, \bibinfo{person}{Ellen Jiang}, \bibinfo{person}{Aaron Donsbach}, \bibinfo{person}{Jeff Gray}, \bibinfo{person}{Alejandra Molina}, \bibinfo{person}{Michael Terry}, {and} \bibinfo{person}{Carrie~J Cai}.} \bibinfo{year}{2022}\natexlab{a}.
\newblock \showarticletitle{Promptchainer: Chaining large language model prompts through visual programming}. In \bibinfo{booktitle}{\emph{CHI Conference on Human Factors in Computing Systems Extended Abstracts}}. \bibinfo{pages}{1--10}.
\newblock


\bibitem[Wu et~al\mbox{.}(2022b)]%
        {wu2022ai}
\bibfield{author}{\bibinfo{person}{Tongshuang Wu}, \bibinfo{person}{Michael Terry}, {and} \bibinfo{person}{Carrie~Jun Cai}.} \bibinfo{year}{2022}\natexlab{b}.
\newblock \showarticletitle{Ai chains: Transparent and controllable human-ai interaction by chaining large language model prompts}. In \bibinfo{booktitle}{\emph{Proceedings of the 2022 CHI conference on human factors in computing systems}}. \bibinfo{pages}{1--22}.
\newblock


\bibitem[Xia et~al\mbox{.}(2016)]%
        {xia2016object}
\bibfield{author}{\bibinfo{person}{Haijun Xia}, \bibinfo{person}{Bruno Araujo}, \bibinfo{person}{Tovi Grossman}, {and} \bibinfo{person}{Daniel Wigdor}.} \bibinfo{year}{2016}\natexlab{}.
\newblock \showarticletitle{Object-oriented drawing}. In \bibinfo{booktitle}{\emph{Proceedings of the 2016 CHI Conference on Human Factors in Computing Systems}}. \bibinfo{pages}{4610--4621}.
\newblock


\bibitem[Xiao et~al\mbox{.}(2024)]%
        {xiao2024typedance}
\bibfield{author}{\bibinfo{person}{Shishi Xiao}, \bibinfo{person}{Liangwei Wang}, \bibinfo{person}{Xiaojuan Ma}, {and} \bibinfo{person}{Wei Zeng}.} \bibinfo{year}{2024}\natexlab{}.
\newblock \showarticletitle{TypeDance: Creating semantic typographic logos from image through personalized generation}. In \bibinfo{booktitle}{\emph{Proceedings of the CHI Conference on Human Factors in Computing Systems}}. \bibinfo{pages}{1--18}.
\newblock


\bibitem[Xing et~al\mbox{.}(2016)]%
        {xing2016energy}
\bibfield{author}{\bibinfo{person}{Jun Xing}, \bibinfo{person}{Rubaiat~Habib Kazi}, \bibinfo{person}{Tovi Grossman}, \bibinfo{person}{Li-Yi Wei}, \bibinfo{person}{Jos Stam}, {and} \bibinfo{person}{George Fitzmaurice}.} \bibinfo{year}{2016}\natexlab{}.
\newblock \showarticletitle{Energy-brushes: Interactive tools for illustrating stylized elemental dynamics}. In \bibinfo{booktitle}{\emph{Proceedings of the 29th Annual Symposium on User Interface Software and Technology}}. \bibinfo{pages}{755--766}.
\newblock


\bibitem[Yan et~al\mbox{.}(2023)]%
        {yan2023xcreation}
\bibfield{author}{\bibinfo{person}{Zihan Yan}, \bibinfo{person}{Chunxu Yang}, \bibinfo{person}{Qihao Liang}, {and} \bibinfo{person}{Xiang'Anthony' Chen}.} \bibinfo{year}{2023}\natexlab{}.
\newblock \showarticletitle{XCreation: A Graph-based Crossmodal Generative Creativity Support Tool}. In \bibinfo{booktitle}{\emph{Proceedings of the 36th Annual ACM Symposium on User Interface Software and Technology}}. \bibinfo{pages}{1--15}.
\newblock


\bibitem[Yang et~al\mbox{.}(2024)]%
        {yang2024gpt}
\bibfield{author}{\bibinfo{person}{Daijin Yang}, \bibinfo{person}{Erica Kleinman}, {and} \bibinfo{person}{Casper Harteveld}.} \bibinfo{year}{2024}\natexlab{}.
\newblock \showarticletitle{GPT for Games: A Scoping Review (2020-2023)}.
\newblock \bibinfo{journal}{\emph{arXiv preprint arXiv:2404.17794}} (\bibinfo{year}{2024}).
\newblock


\bibitem[Yao et~al\mbox{.}(2024)]%
        {yao2024tree}
\bibfield{author}{\bibinfo{person}{Shunyu Yao}, \bibinfo{person}{Dian Yu}, \bibinfo{person}{Jeffrey Zhao}, \bibinfo{person}{Izhak Shafran}, \bibinfo{person}{Tom Griffiths}, \bibinfo{person}{Yuan Cao}, {and} \bibinfo{person}{Karthik Narasimhan}.} \bibinfo{year}{2024}\natexlab{}.
\newblock \showarticletitle{Tree of thoughts: Deliberate problem solving with large language models}.
\newblock \bibinfo{journal}{\emph{Advances in Neural Information Processing Systems}}  \bibinfo{volume}{36} (\bibinfo{year}{2024}).
\newblock


\bibitem[Ye et~al\mbox{.}(2024)]%
        {ye2024prointerar}
\bibfield{author}{\bibinfo{person}{Hui Ye}, \bibinfo{person}{Jiaye Leng}, \bibinfo{person}{Pengfei Xu}, \bibinfo{person}{Karan Singh}, {and} \bibinfo{person}{Hongbo Fu}.} \bibinfo{year}{2024}\natexlab{}.
\newblock \showarticletitle{ProInterAR: A Visual Programming Platform for Creating Immersive AR Interactions}. In \bibinfo{booktitle}{\emph{Proceedings of the CHI Conference on Human Factors in Computing Systems}}. \bibinfo{pages}{1--15}.
\newblock


\bibitem[Yen et~al\mbox{.}(2023)]%
        {yen2023coladder}
\bibfield{author}{\bibinfo{person}{Ryan Yen}, \bibinfo{person}{Jiawen Zhu}, \bibinfo{person}{Sangho Suh}, \bibinfo{person}{Haijun Xia}, {and} \bibinfo{person}{Jian Zhao}.} \bibinfo{year}{2023}\natexlab{}.
\newblock \showarticletitle{Coladder: Supporting programmers with hierarchical code generation in multi-level abstraction}.
\newblock \bibinfo{journal}{\emph{arXiv preprint arXiv:2310.08699}} (\bibinfo{year}{2023}).
\newblock


\bibitem[Yigitbas et~al\mbox{.}(2023)]%
        {yigitbas2023end}
\bibfield{author}{\bibinfo{person}{Enes Yigitbas}, \bibinfo{person}{Jonas Klauke}, \bibinfo{person}{Sebastian Gottschalk}, {and} \bibinfo{person}{Gregor Engels}.} \bibinfo{year}{2023}\natexlab{}.
\newblock \showarticletitle{End-user development for interactive web-based virtual reality scenes}.
\newblock \bibinfo{journal}{\emph{Journal of Computer Languages}}  \bibinfo{volume}{74} (\bibinfo{year}{2023}), \bibinfo{pages}{101187}.
\newblock


\bibitem[Zamfirescu-Pereira et~al\mbox{.}(2023)]%
        {zamfirescu2023johnny}
\bibfield{author}{\bibinfo{person}{JD Zamfirescu-Pereira}, \bibinfo{person}{Richmond~Y Wong}, \bibinfo{person}{Bjoern Hartmann}, {and} \bibinfo{person}{Qian Yang}.} \bibinfo{year}{2023}\natexlab{}.
\newblock \showarticletitle{Why Johnny can’t prompt: how non-AI experts try (and fail) to design LLM prompts}. In \bibinfo{booktitle}{\emph{Proceedings of the 2023 CHI Conference on Human Factors in Computing Systems}}. \bibinfo{pages}{1--21}.
\newblock


\bibitem[Zhang and Oney(2020)]%
        {zhang2020flowmatic}
\bibfield{author}{\bibinfo{person}{Lei Zhang} {and} \bibinfo{person}{Steve Oney}.} \bibinfo{year}{2020}\natexlab{}.
\newblock \showarticletitle{Flowmatic: An immersive authoring tool for creating interactive scenes in virtual reality}. In \bibinfo{booktitle}{\emph{Proceedings of the 33rd Annual ACM Symposium on User Interface Software and Technology}}. \bibinfo{pages}{342--353}.
\newblock


\bibitem[Zhang et~al\mbox{.}(2023)]%
        {zhang2023critical}
\bibfield{author}{\bibinfo{person}{Quanjun Zhang}, \bibinfo{person}{Tongke Zhang}, \bibinfo{person}{Juan Zhai}, \bibinfo{person}{Chunrong Fang}, \bibinfo{person}{Bowen Yu}, \bibinfo{person}{Weisong Sun}, {and} \bibinfo{person}{Zhenyu Chen}.} \bibinfo{year}{2023}\natexlab{}.
\newblock \showarticletitle{A critical review of large language model on software engineering: An example from chatgpt and automated program repair}.
\newblock \bibinfo{journal}{\emph{arXiv preprint arXiv:2310.08879}} (\bibinfo{year}{2023}).
\newblock


\end{thebibliography}











\end{document}